\documentclass[a4paper,12pt]{article}
\usepackage{amsmath}
\usepackage{amssymb} 
\pdfoutput = 1

\usepackage{subcaption}
\usepackage{graphicx}

\usepackage[left=2cm, right=2cm, top=2cm]{geometry}

\begin{document}
\begin{titlepage}
\renewcommand{\thefootnote}{\fnsymbol{footnote}}
\begin{flushright}
ICAS 044/19 \\
     \end{flushright}
\par \vspace{10mm}
\begin{center}
{\Large \bf  Electroweak boson production at large transverse  \\[5mm]  momentum  
in double polarized $pp$  scattering  \\[5mm]  at next-to-leading order accuracy}

\end{center}
\par \vspace{2mm}
\begin{center}
{\bf Daniel de Florian and Iv\'an Pedron} \\
\vspace{5mm}
International Center for Advanced Studies (ICAS) and ICIFI, \\ ECyT-UNSAM,
Campus Miguelete, \\ 25 de Mayo y Francia, (1650) Buenos Aires, Argentina

\end{center}


\par \vspace{9mm}
\begin{center} {\large \bf Abstract} \end{center}
\begin{quote}
\pretolerance 10000
We  compute the next-to-leading order corrections to the cross section for the production of electroweak gauge bosons ($Z$, $W^\pm$ and $\gamma^*$) with large transverse momentum in double (longitudinally) polarized hadronic collisions.
The calculation is fully performed in the HVBM scheme within dimensional regularization with a careful treatment of issues arising due to the appearance of $\gamma^5$ in polarization projectors and axial couplings.

We study the phenomenological consequences of the NLO corrections at the level of both the double polarized cross section and the corresponding asymmetries at RHIC kinematics.

\end{quote}

\end{titlepage}

\setcounter{footnote}{1}
\renewcommand{\thefootnote}{\fnsymbol{footnote}}


\section{Introduction}

Over the last decades, our understanding of hadron structure has remarkably improved 
thanks to impressive experimental and theoretical progress. 
In particular,  there is currently much  activity aiming at further unravelling 
the nucleon's spin structure. 

As is well known, the total quark and 
anti-quark  spin contribution to the nucleon spin 
was found to be only about $\sim 25 \%$\cite{deFlorian:2014yva,Nocera:2014gqa,Sato:2016tuz},  information that has been mostly provided
by  Deep-Inelastic scattering (DIS) data.

One emphasis is on the determination of the 
spin-dependent gluon distribution, $\Delta g$,
of the nucleon, which ultimately would give the gluon  contribution to
the nucleon spin. 
Clear evidence for a non-vanishing polarization of gluons was found in the region of momentum fraction and at the scales mostly probed by the BNL Relativistic Heavy Ion Collider (RHIC) data \cite{deFlorian:2014yva}. But from the available results it is not possible yet to set a precise value on the total gluon contribution to the proton spin and  to the flavor decomposition of the quarks \footnote{See \cite{deFlorian:2019egz} for an analysis of the proton spin budget at three loops.}.

Spin asymmetries in 
high-energy $pp$ scattering can be particularly sensitive to $\Delta g$, 
for processes where gluons in the initial state contribute already at 
the lowest order of perturbation theory. Furthermore, the Drell-Yan process, the production
of vector bosons \cite{Adare:2018csm,Adam:2018bam}, has been shown to be very relevant to help for quark flavor separation,
 since quarks couple differently to both $W^{\pm}$ and $Z$ bosons \cite{deFlorian:2010aa}.

Therefore, one particularly interesting process that satisfies both requirements 
is the single-inclusive production of large transverse-momentum 
($q_T$) vector bosons, $pp \to V X$, that can help to provide information on both $\Delta g$ and the flavour decomposition of the quark contribution.

In order to make reliable quantitative predictions for a high-energy process, it is crucial to determine the next-to-leading order (NLO) QCD corrections to the Born approximation. In general, the key issue here is to check the perturbative stability of the process considered, i.e. to examine to what extent  the NLO corrections affect the cross sections and  spin asymmetries relevant for experimental measurements. Only if the corrections are under control can a process that shows good sensitivity to, say, $\Delta g$ at the lowest order be regarded as a genuine probe of the polarized gluon distribution and be reliably used to extract it from future data.

In the particular case of (large transverse momentum) gauge boson production another  perturbative issues appear due to the existence of two physical scales, the transverse momentum $q_T$ and the `mass' $Q$ of the boson (or its virtuality, specially in the case of  photon production).

In the large-$q_T$ region ($q_T\sim Q$), where the transverse momentum is of the order of the vector boson mass, the QCD perturbative series is controlled by a small expansion parameter, $\alpha_s(Q)$, and  calculations based on the truncation of the perturbative series at a fixed order in $\alpha_s$ 
are theoretically justified. In this region, the QCD radiative corrections for the unpolarized cross section are known up to the next-to-leading order (NLO) \cite{Ellis:1981hk,Arnold:1988dp,Gonsalves:1989ar} in an  analytical form and next-to-next-to-leading order (NNLO) corrections were recently obtained in numerical implementations in \cite{Ridder:2016nkl}--\cite{Gehrmann-DeRidder:2017mvr}. In the polarized case, a first attempt to achieve the cross section at NLO accuracy for virtual photon production was presented in \cite{Chang:1997ik}, where only the non-singlet contribution was obtained. The full NLO result for the double polarized cross section, including the relevant gluon initiated channels, was missing so far.

Nonetheless the bulk of the vector boson events (particularly at RHIC with a center-of-mass energy of $\sqrt{S}=510$ GeV) is produced in the small-$q_T$ region ($q_T\ll Q$), where the convergence of the fixed-order expansion is spoiled by the presence of large logarithmic terms, $\alpha_s^n\ln^m (Q^2/q_T^2)$. To obtain reliable predictions, these logarithmically-enhanced terms 
have to be systematically resummed to all perturbative orders (see e.g.\cite{Catani:2015vma},\cite{Bizon:2019zgf} and references therein). But even in that case, the fixed-order calculation becomes essential in order to perform the proper matching with the resummed contribution, affecting the full result even at rather small transverse momentum. Furthermore, since the small-$q_T$ region is mostly affected by soft gluon emission, which is independent on the polarization of the emitting parton, the resummation typically affects both polarized and unpolarized cross section in a rather similar way, rendering a very small effect at the level of asymmetries. 

Therefore, counting with the NLO corrections for polarized hadronic collisions becomes fundamental in both kinematical regimes in order to understand the data produced at RHIC and to extract the corresponding information in terms of polarized partonic distributions. In this paper we compute the NLO corrections to the cross section for the production of gauge bosons ($Z$, $W^\pm$ and $\gamma^*$) with large transverse momentum in double (longitudinally) polarized hadronic collisions.

The paper is organized as follows. In Sect.~\ref{sect1} we briefly review the procedure to compute the NLO corrections with emphasis on the issues arising due to the treatment of $\gamma^5$ in dimensional regularization. In Sect.~\ref{sect2} we present the analytical results for the NLO corrections for gauge boson  production (some of the lengthly functions are given in the Appendix). In Sect.~\ref{sect3} we study the phenomenological impact of the NLO corrections at RHIC kinematics, including the measurable double longitudinal asymmetries. Finally, in Sect.~\ref{sec:summa} we summarize our results.


\section{Cross Section Calculation}
\label{sect1}

\subsection{Inclusive cross section}

We calculate the inclusive cross section of electroweak boson production in the framework of perturbative QCD, considering the case of both polarized and unpolarized initial hadrons. The process is described as

\begin{equation}
 h_1(P_1)+h_2(P_2) \rightarrow V(Q)+X,
\end{equation}

\noindent where $h_i$, $i=1,2$ are the polarized/unpolarized initial hadrons with momenta $P_i$, and $V$ is either a $W^{\pm}$, $Z^0$ or virtual photon $\gamma^*$ with $Q^2=M_V^2$, energy $E_Q=Q^0$ and large transverse momentum $q_T$ with respect to the collision axis. The unpolarized $\sigma$ and polarized $\Delta \sigma$ cross sections are obtained from the sum or difference of helicity-dependent cross sections given by

\begin{equation}
 \sigma = \frac{1}{4} \left( \sigma^{++} + \sigma^{+-} + \sigma^{-+} + \sigma^{--} \right),
\end{equation}

\noindent for the unpolarized case, and

\begin{equation}
 \Delta \sigma = \frac{1}{4} \left( \sigma^{++} - \sigma^{+-} - \sigma^{-+} + \sigma^{--} \right),
\end{equation}

\noindent for the polarized one. Here the superindices $+,-$ denote the helicities of the two incoming hadrons.

The inclusive (polarized) unpolarized cross section can be expressed as

\begin{equation}
 E_Q \frac{d (\Delta) \sigma}{d^3 Q} = \sum_{a,b} \int^1_0 dx_a \ dx_b \ (\Delta)f_a^{h_1}(x_a,\mu_F^2) \ (\Delta)f_b^{h_2}(x_b,\mu_F^2) \ E_Q \frac{d (\Delta) \hat{\sigma}^{a,b}}{d^3 Q}(p_a, p_b, \mu_F^2).
\end{equation}

\noindent Here $(\Delta)f^h_a(x, \mu_F^2)$ is the (polarized) unpolarized parton distribution function (PDF) of parton $a$ with momentum fraction $x$ in hadron $h$, probed at the scale $\mu_F^2$, which are given by the combinations

\begin{equation}
 f_a(x,\mu_F^2)=f^{+}_a(x,\mu_F^2)+f_a^{-}(x,\mu_F^2), \quad \Delta f_a(x,\mu_F^2)=f_a^{+}(x,\mu_F^2)-f_a^{-}(x,\mu_F^2).
\end{equation}

\noindent In this case, the superindex $+,-$ denotes de helicity orientation of the parton $a$ with respect to the helicity of the parent hadron. The perturbative (polarized) unpolarized cross section $(\Delta) \hat{\sigma}^{a,b}(p_a, p_b, \mu_F^2)$ corresponds to the hard-scattering partonic process

\begin{equation}
 a(p_a)+b(p_b) \rightarrow V(Q)+X,
\end{equation}

\noindent where $p_i$ represents the parton momentum. Here collinear singularities due to the radiation of massless partons are factorized out at the scale $\mu_F^2$ and included in the scale dependent (polarized) unpolarized PDFs $(\Delta)f^h_a(x, \mu_F^2)$.

The partonic momenta can be described in terms of the hadronic ones using the relation $p_i=x_i P_i$. As it is customary, we introduce the Mandelstam variables for both the hadronic and partonic levels
\begin{equation}
    \begin{aligned}[c]
        S \equiv &(P_1+P_2)^2, \quad T \equiv (P_1-Q)^2, \quad U \equiv (P_2-Q)^2, \\
        s \equiv &(p_a+p_b)^2, \quad t \equiv (p_a-Q)^2, \quad u \equiv (p_b-Q)^2, \\
        S_{23}& \equiv S+T+U-Q^2, \quad s_{23} \equiv s+t+u-Q^2.
    \end{aligned}
\end{equation}

\noindent $S$ and $s$ are the hadronic and partonic invariant center-of-mass energies squared of the colliding system, respectively, while $s_{23}$ is the invariant mass squared of the system recoiling against the boson $V$. The partonic cross section $\hat{\sigma}^{a,b}$ has a singular behaviour in the limit $s_{23} \rightarrow 0$ related to the cancellation of singularities due to soft gluon emission in the recoiling system and virtual gluon infrared singularities. However, the cross section is actually integrable and the  $1/s_{23}$ terms can be dealt with the following change of variables in the momentum fraction integration

\begin{equation}
    \int^1_0 \int^1_0 dx_1 \ dx_2 \ \theta(s_{23}) \ \theta(p_1^0+p_2^0-E_Q)=\int^1_B dx_1 \ \frac{1}{x_1 \ S + U - Q^2} \ \int^A_0 ds_{23},
\label{eq_ints23}
\end{equation}

\noindent where the new integration limits are given by
\begin{equation}
    A=U+x_1 \ (S_{23}-U), \quad B=-\frac{U}{S_{23}+U}.
\end{equation}

\subsection{Perturbative calculation techniques}

\begin{figure}
\centering
  \includegraphics[width=0.4\linewidth]{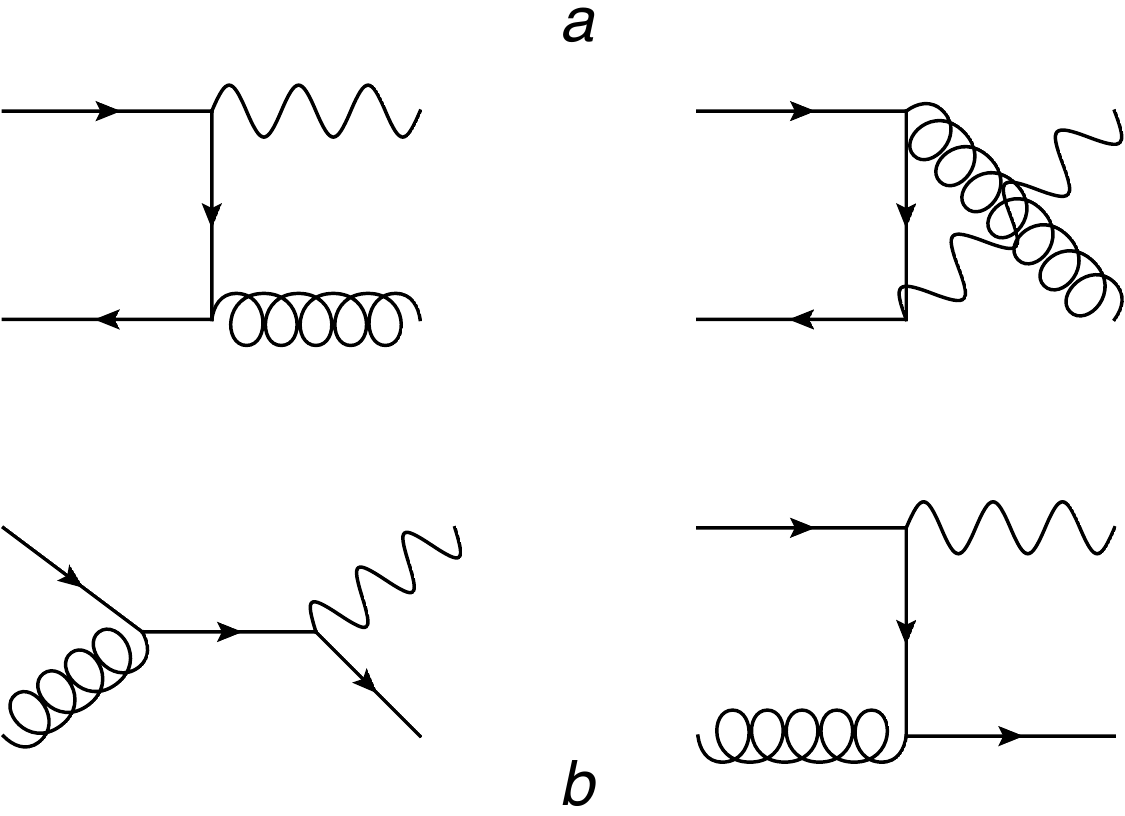}
  \caption{Born diagrams of the two processes at LO.}
  \label{fig_FeynBorn}
\end{figure}

The analytical procedures to compute the perturbative cross section are already well established. At the lowest order (LO) only two channels contribute, as depicted in Fig.\ref{fig_FeynBorn}: the annihilation process $q \overline{q} \rightarrow V g$ (a) and the Compton process $q g \rightarrow V q$ (b). At the following order (NLO) virtual and real contributions must be considered. In this case, there are four contributing channels: the process initiated by a quark and an antiquark $q \overline{q} \rightarrow V (q \overline{q}, \ q' \overline{q}', \ gg)$, the one initiated by two quarks $q q \rightarrow V q q$, the process initiated by a single gluon $q g \rightarrow V q g$ and the two gluon process $g g \rightarrow V q \overline{q}$. 

In all processes we evaluate the Feynman diagrams, performing the Dirac traces with the TRACER package \cite{Jamin:1991dp}, and we integrate over the virtual loop and real emission momenta. Ghost graphs are taken into account to simplify the polarization sums of external gluons. Real contributions are computed using traditional partial fractioning and master integrals, while virtual loop contributions are calculated using the FeynCalc \cite{Shtabovenko:2016sxi} and Feynhelpers \cite{Shtabovenko:2016whf} packages. Dimensional regularization is used to regulate infrared and ultraviolet singularities, working in a space-time of $d=4-2 \epsilon $ dimensions. As it will be addressed in the following subsections, dimensional regularization requires a special treatment for the Levi-Civita tensor $\epsilon^{\mu \nu \rho \sigma}$ and the Dirac matrix $\gamma^5$, which is present in both the electroweak couplings and quirality projectors. The factorization of the mass singularities in the polarized case is also affected by the scheme used to deal with $\gamma^5$.

\subsubsection{Treatment of the matrix $\gamma_5$ and axial couplings}

While dimensional regularization involves working in $d$-dimensional space-time, the $\gamma^5$ matrix and the $\epsilon^{\mu \nu \rho \sigma}$ tensor are only well defined in the four-dimensional space-time. A consistent way to treat $\gamma^5$ and $\epsilon^{\mu \nu \rho \sigma}$ in $d$-dimensions is the HVBM scheme \cite{tHooft:1972tcz,Breitenlohner:1977hr}, which splits the $d$-dimensional Minkowsky space into the usual four-dimensional one and a $(d-4)$-dimensional subspace where, for instance, the $(d-4)$-dimensional part of the $\gamma^{\mu}$ matrices, represented as $\hat{\gamma}^{\mu}$, commutes with the strictly four-dimensional $\gamma^5$. Calculations in the HVBM scheme are algebraically more involved because for unobserved momenta $p$ and $\gamma^{\mu}$ matrices one needs to take into account both their four-dimensional part ($\tilde{p}$, $\tilde{\gamma^{\mu}}$) and their $(d-4)$-dimensional one ($\hat{p}$, $\hat{\gamma^{\mu}}$), which follow different algebraic treatment. Integration over `hat' momenta $\hat{p}$ in the real contributions is performed as established in  Ref.\cite{Gordon:1993qc}.

One of the first issues related to the HVBM scheme is the definition of the electroweak vertices. The standard definition of the boson vertices, taking into account the flavours $f_1$ and $f_2$ of the involved quarks, can be expressed as

\begin{equation}
    -i e \gamma_{\mu} \left( L_{f_2 f_1} \frac{1-\gamma_5}{2} + R_{f_2 f_1} \frac{1+\gamma_5}{2} \right),
\label{eq_vert4D}
\end{equation}

\noindent where the left and right-handed coupling are given by the following expressions, depending on the type of boson

\begin{equation}
    \begin{aligned}[c]
        W^{-}: & \quad L_{f_2 f_1}=\frac{1}{\sqrt{2} \sin \theta_W}(\tau_{+})_{f_2 f_1} U_{f_2 f_1}, \quad R_{f_2 f_1}=0, \\
        W^{+}: & \quad L_{f_2 f_1}=\frac{1}{\sqrt{2} \sin \theta_W}(\tau_{-})_{f_2 f_1} U^{\dagger}_{f_2 f_1}, \quad R_{f_2 f_1}=0 \\
        Z^{0}: & \quad L_{f_2 f_1}=\frac{1}{\sin (2 \theta_W)}(\tau_3)_{f_2 f_1}- \delta_{f_2 f_1} e_{f_1} \tan \theta_W, \quad R_{f_2 f_1}=-\delta_{f_2 f_1} e_{f_1} \tan \theta_W, \\
        \gamma: & \quad L_{f_2 f_1}=\delta_{f_2 f_1} e_{f_1}, \quad R_{f_2 f_1}=\delta_{f_2 f_1} e_{f_1}. \\
    \end{aligned}
\end{equation}

\noindent Here $\theta_W$ is the electroweak mixing angle, $\tau_{\pm}=(\tau_1 \pm i \tau_2)/2$ and $\tau_3$ are the weak isospin Pauli matrices, $U$ is the CKM mixing matrix, and $e_f$ is the electric charge of the corresponding quark ($e_f=\frac{2}{3}$ for $u,c,t$ and $e_f=-\frac{1}{3}$ for $d,s,b$). However, since the commutation properties of the combination $\gamma_{\mu} \gamma_{5}$ depends on whether $\mu \leq 3$ or $\mu > 3$, a symmetric definition of the vertices is needed, which corresponds to the replacement $\gamma_{\mu}(1-\gamma_5) \rightarrow \frac{1}{2}(1+\gamma_5)\gamma_{\mu}(1-\gamma_5)$ \cite{Korner:1989is,Buras:1989xd}. After this substitution, the vertex expression of Eq.(\ref{eq_vert4D}) in $d$-dimensions can be rewritten as  

\begin{equation}
    -i e \tilde{\gamma}_{\mu} \left( L_{f_2 f_1} \frac{1-\gamma_5}{2} + R_{f_2 f_1} \frac{1+\gamma_5}{2} \right), 
\end{equation}

\noindent where the $(d-4)$-dimensional part of $\gamma_{\mu}$ cancels within the left and right-handed terms. 

In the HVBM scheme the presence of $\gamma_5$ in both the symmetrized vertex and the quirality projectors over initial partons lead to anomalous terms of order $\mathcal{O}(d-4)$, which in combination with divergent terms can result in spurious finite contributions. Infrared anomalous terms cancel out between virtual and real contributions \cite{Korner:1985uj}, but in the case of the renormalization of ultraviolet divergencies and mass factorization appropriate additional counterterms are required.

\subsubsection{Renormalization and the axial vertex anomaly}

For the renormalization of the virtual contributions to the NLO we adopt de usual $\overline{MS}$ scheme. In this scheme the running strong coupling constant $\alpha_s(\mu^2)$ at a scale $\mu^2$ satisfies the corresponding renormalization-group equation

\begin{equation}
    \mu^2 \frac{d}{d \mu^2} \alpha_s (\mu^2) =-\alpha_s (\mu^2) \left[ \beta_0 \frac{\alpha_s (\mu^2)}{2 \pi} + \beta_1 \left( \frac{\alpha_s (\mu^2)}{2 \pi} \right)^2+ \mathcal{O}(\alpha_s^3) \right], 
\end{equation}

\noindent with the $\beta_i$ factors given by

\begin{equation}
    \begin{aligned}[c]
        \beta_0 &= \frac{11}{6} C_A-\frac{1}{3} \sum_f (1), \\
	\beta_1 &= \frac{17}{6} C_A^2-\left(\frac{5}{6} C_A + \frac{1}{2} C_F \right) \sum_f (1),
    \end{aligned}
\end{equation}

\noindent where the sum $\sum_f$ runs over all the flavours below the mass threshold.

However, within the HVBM scheme, since $\gamma^5$ no longer anticommutes with $\gamma^{\mu}$ an additional finite renormalization of the axial quark current needs to be taken into account at NLO \cite{Larin:1993tq}. The renormalization constant related to the axial current $Z_5$ is no longer equal to one. The corresponding one loop expression is

\begin{equation}
    Z_5 =1-\frac{\alpha_s}{2 \pi} \ 2 C_F+\mathcal{O}(\alpha_s^2). 
\end{equation}

\noindent This term gives rise to the following additional finite counterterms that have to be added to the $q\overline{q}$ and $qg$ channels in order cancel the spurious axial terms appearing in the virtual contributions
\begin{equation}
    \begin{aligned}[c]
      (\Delta)  C^{q\overline{q}}=&-\frac{\alpha_s}{2 \pi} \left( 2 \ C_F\right) \frac{d (\Delta) \sigma^{axial}_{q\overline{q}}}{d^3 Q}, \\
	(\Delta)C^{qg}=&-\frac{\alpha_s}{2 \pi} \left( 2 \ C_F\right) \frac{d (\Delta) \sigma^{axial}_{qg}}{d^3 Q},
    \end{aligned}
\label{eq_axrenct}
\end{equation}

\noindent where $d (\Delta) \sigma^{axial}_{ij}/d^3 Q$ is the axial part of the (polarized) unpolarized Born cross section corresponding to either the $q \overline{q} \rightarrow V g$ process or the  $q g \rightarrow V q$ Compton process. This axial cross section is obtained by considering only the terms depending on the axial coupling to the boson $V$. 

\subsubsection{Factorization}

Mass singularities appear due to the collinear emission of massless partons from one of the incoming partons. These singular terms are detached at a scale $\mu_F^2$ to be incorporated into the (polarized) unpolarized distribution functions $(\Delta)f_a^h(x,\mu_F^2)$. For the unpolarized case, the results are calculated in the usual $\overline{MS}$ scheme. However, in the polarized case we use the conventional variation of the $\overline{MS}$ scheme which takes into account some helicity-conservation violations that arise in the HVBM scheme, particularly in the quark-antiquark annihilation process.

\begin{figure}
\centering
  \includegraphics[width=0.4\linewidth]{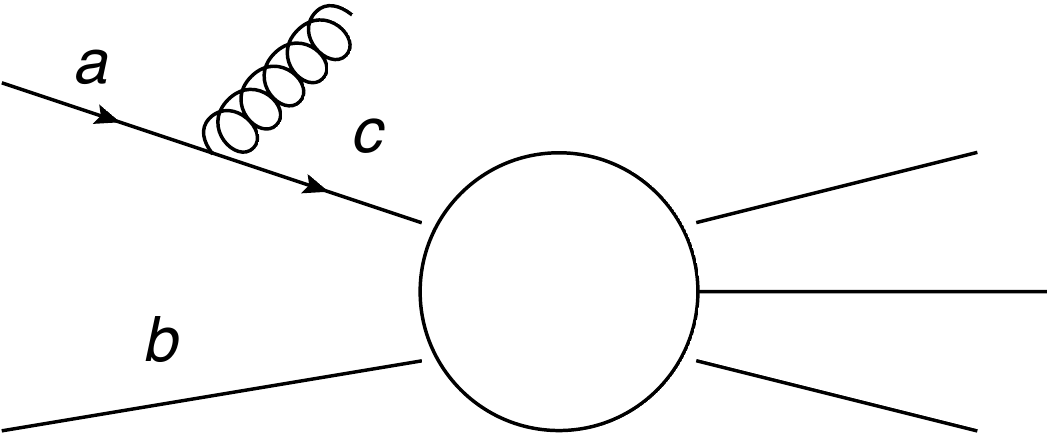}
  \caption{Schematic of the parton emission processes.}
  \label{fig_fact}
\end{figure}

The factorization subtractions \footnote{Subtractions that need to be applied to the corresponding partonic channel.} due to collinear emission from an initial parton $a$, as illustrated in Fig.\ref{fig_fact} for gluon emission, are given by the following convolution

\begin{equation}
    d (\Delta) \hat{\sigma}^{fact}_{a b \rightarrow V X}=\frac{\alpha_s}{2 \pi} \left\{ \left[ -\frac{1}{\hat{\epsilon}} (\Delta) P_{ca}(z) \left(\frac{\mu^2}{\mu_F^2} \right)^{\epsilon}+ (\Delta) f_{ca} (z)\right] \circledast d (\Delta) \sigma_{cb \rightarrow V X} \right\}, 
\end{equation}

\noindent where $\mu$ is the scale of dimensional regularization and $\mu_F$ represents the factorization scale. Here $1/\hat{\epsilon}=1/\epsilon-\gamma_E+\ln 4 \pi$ as in the $\overline{MS}$ scheme, $(\Delta) P_{ca}(z)$ is the (polarized) unpolarized Altarelli-Parisi splitting function of a parton $a$ emitting a parton $c$ carrying momentum fraction $z$, $d (\Delta) \sigma_{cb \rightarrow V X}$ is the (polarized) unpolarized $d$-dimensional Born cross section of the partonic process $c b \rightarrow V X$, and $(\Delta) f_{ab}(z)$ represents the freedom on subtracting additional finite terms. In the unpolarized case we rely on the usual $\overline{MS}$ scheme, which corresponds to $f_{ij}(z)=0$, but in the polarized case when the subtraction involves emission of quarks by other quarks, when working with the $\Delta P_{qq}(z)$ kernel, we use $\Delta f_{qq}(z)=-4 \ C_F (1-z)$. This is the particular scheme is used in the definition of the polarized parton distribution sets \cite{deFlorian:2009vb}. 

\section{NLO corrections}
\label{sect2}

In this section we present the NLO contributions to the polarized cross section of electroweak boson production at large transverse momentum in hadronic collisions. For the sake of completeness we also provide the unpolarized results, which are in agreement with those of Gonsalves et al. \cite{Gonsalves:1989ar}, with the distinction that we corroborated the calculation fully within the consistent HVBM scheme for the axial coupling. The polarized cross section for the virtual photon in the non-singlet case partially agrees with the previous results of Ref.\cite{Chang:1997ik} due to an overlooked sign in one of the interferences.

\subsection{Quark-antiquark annihilation and scattering}

The formulas of the inclusive cross section are presented in the same way as in Ref.\cite{Gonsalves:1989ar}, with the expression separated in contributions with different axial-vector and/or flavour structure. The quark-antiquark cross section formula is given by

\begingroup
\small
\begin{flalign}
 E_Q \frac{d (\Delta) \sigma^{q\overline{q}}}{d^3 Q}=&\frac{\alpha \ \alpha_s^{\overline{MS}}(\mu^2) \ C_F}{s \ N_c} \Bigg\{ \delta (s_{23}) \ (\Delta) B_{q\overline{q}}(s,t,u,Q^2) \left( |L_{21}|^2+|R_{21}|^2 \right)+\frac{\alpha_s^{\overline{MS}}(\mu^2)}{2 \pi} \notag & \\
 & \times \Big\{ \Big[ \delta (s_{23}) \big[ (\Delta) V^{(3)}_{q\overline{q}}(s,t,u,Q^2) + (\Delta) V^{(1)}_{q\overline{q}}(s,t,u,Q^2) \sum_f (1) \big] \notag \\
 &\quad + (\Delta) G_{q\overline{q}}(s,t,u,Q^2) + (\Delta) F_{aa}(s,t,u,Q^2) \sum_f (1) \Big] \left( |L_{21}|^2+|R_{21}|^2 \right) \notag \\
 &\quad + \Big[ \delta (s_{23}) \ (\Delta) V^{(2)}_{q\overline{q}}(s,t,u,Q^2) + (\Delta) F_{ab}(s,t,u,Q^2) \Big] \delta_{12} \left(L_{11}-R_{11}\right) \sum_f \left(L_{ff}-R_{ff}\right) \notag \\
 &\quad + (\Delta) F_{bb}(s,t,u,Q^2) \delta_{12} \sum_f \sum_{f'} \left( |L_{ff'}|^2+|R_{ff'}|^2 \right) \notag \\
 &\quad +\Big[ (\Delta) F_{ac}(s,t,u,Q^2) + (\Delta) F_{ad}(s,t,u,Q^2) \Big] \left( |L_{21}|^2+|R_{21}|^2 \right) \notag \\
 &\quad +\Big[ \delta_{12} (\Delta) F_{bc}(s,t,u,Q^2) + (\Delta) F_{cc}(s,t,u,Q^2) \Big] \sum_f \left( |L_{f1}|^2+|R_{f1}|^2 \right) \notag \\
 &\quad +\Big[ \delta_{12} (\Delta) F_{bd}(s,t,u,Q^2) + (\Delta) F_{dd}(s,t,u,Q^2) \Big] \sum_f \left( |L_{2f}|^2 +|R_{2f}|^2 \right) \notag \\
 &\quad + (\Delta) F^{LL}_{cd}(s,t,u,Q^2) \left( L_{11} L_{22}+R_{11} R_{22} \right) \notag \\
 &\quad + (\Delta) F^{LR}_{cd}(s,t,u,Q^2) \left( L_{11} R_{22}+R_{11} L_{22} \right)\Big\} \Bigg\}.
\label{eq_masterqqb}
\end{flalign}
\endgroup

\noindent Here the $(\Delta) B_{q\overline{q}}(s,t,u,Q^2)$ function corresponds to the (polarized) unpolarized Born diagrams contributions \footnote[1]{The axial Born cross section needed for the axial renormalization counterterm in Eq.(\ref{eq_axrenct}) is given by the expression $E_Q \frac{d (\Delta) \sigma^{axial}_{q\overline{q}}}{d^3 Q}=\frac{\alpha \ \alpha_s^{\overline{MS}}(\mu^2) \ C_F}{s \ N_c} (\Delta) B_{q\overline{q}}(s,t,u,Q^2) \frac{\left( L_{21}-R_{21} \right)^2}{2}$. The expression is analogous in the $qg$ case.}

\small
\begin{equation}
 \Delta B_{q\overline{q}}(s,t,u,Q^2)=-\frac{t^2+u^2+2 s Q^2}{t u}, \quad B_{q\overline{q}}(s,t,u,Q^2)=-\Delta B_{q\overline{q}}(s,t,u,Q^2).
\label{eq_Bornqqb}
\end{equation}
\normalsize

\begin{figure}
\centering
  \includegraphics[width=0.6\linewidth]{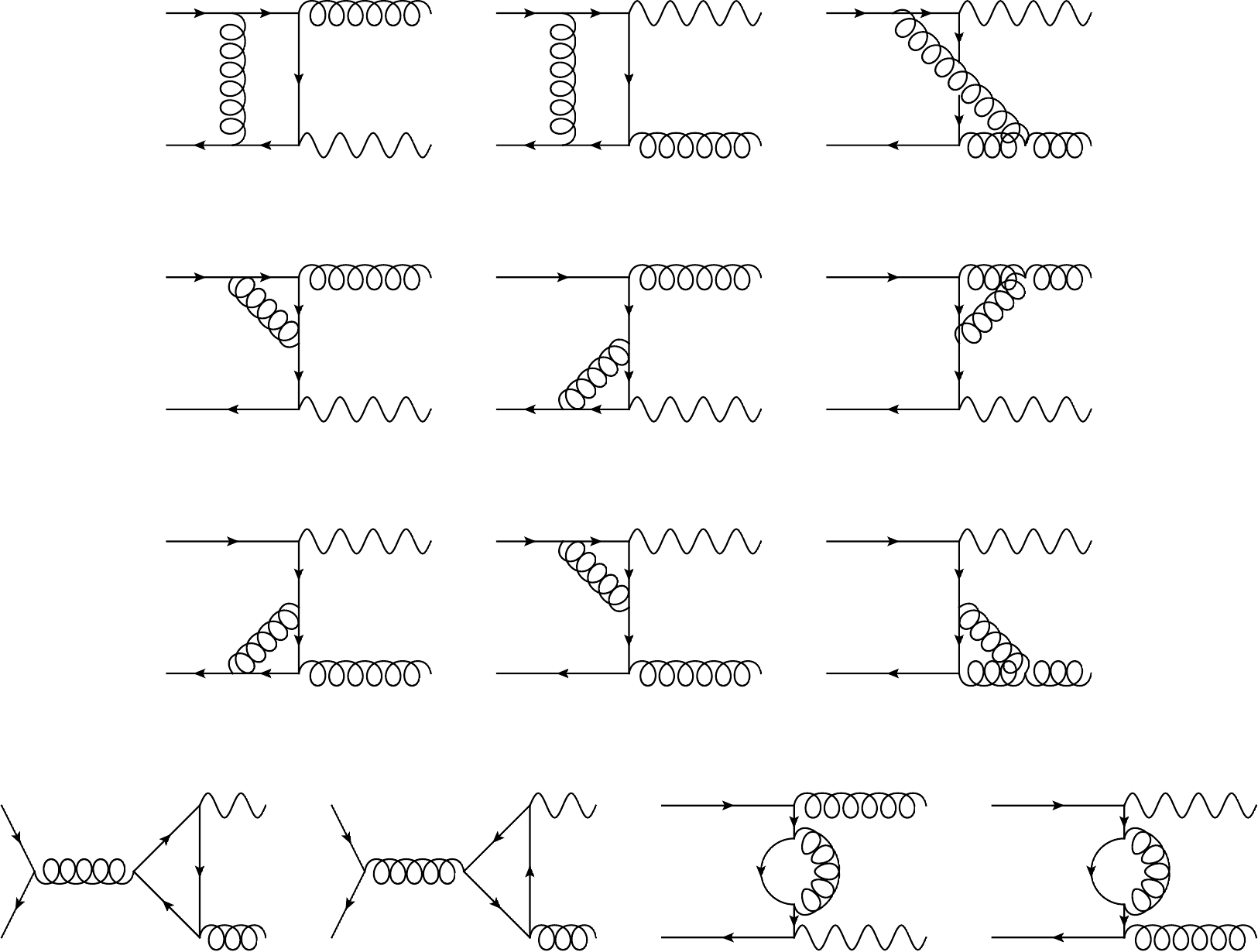}
  \caption{Diagrams corresponding to the virtual corrections to the Born $q \overline{q} \rightarrow V g$ process. The $V$ functions in the cross section formula correspond to these diagrams.}
  \label{fig_FeynVirt}
\end{figure}

\begin{figure}
\centering
  \centering
  \includegraphics[width=0.6\linewidth]{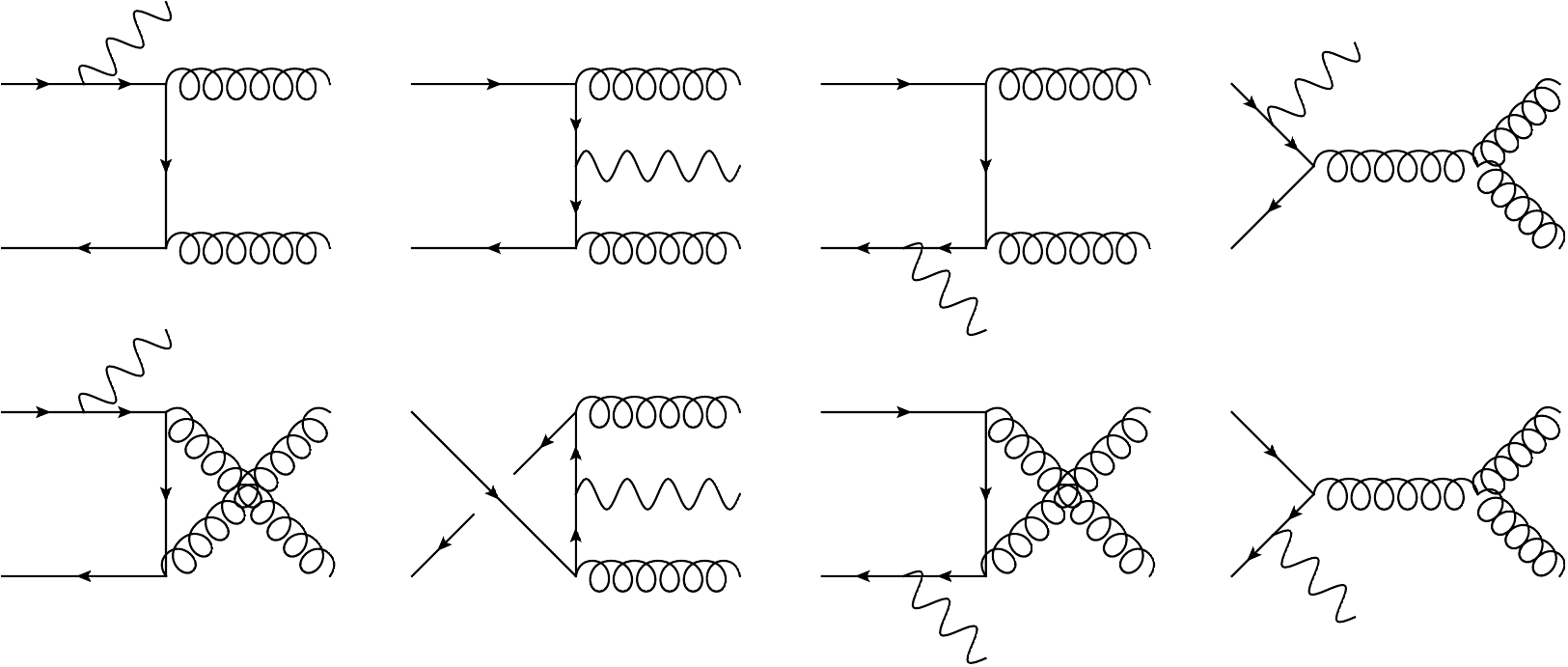}
  \caption{Diagrams which contribute to the annihilation process $q \overline{q} \rightarrow V g g$. Ghost graphs are not depicted. The $G$ functions in the cross section formula correspond to these diagrams.}
  \label{fig_FeynqqGG}
\end{figure}

\begin{figure}
  \centering
  \includegraphics[width=0.6\linewidth]{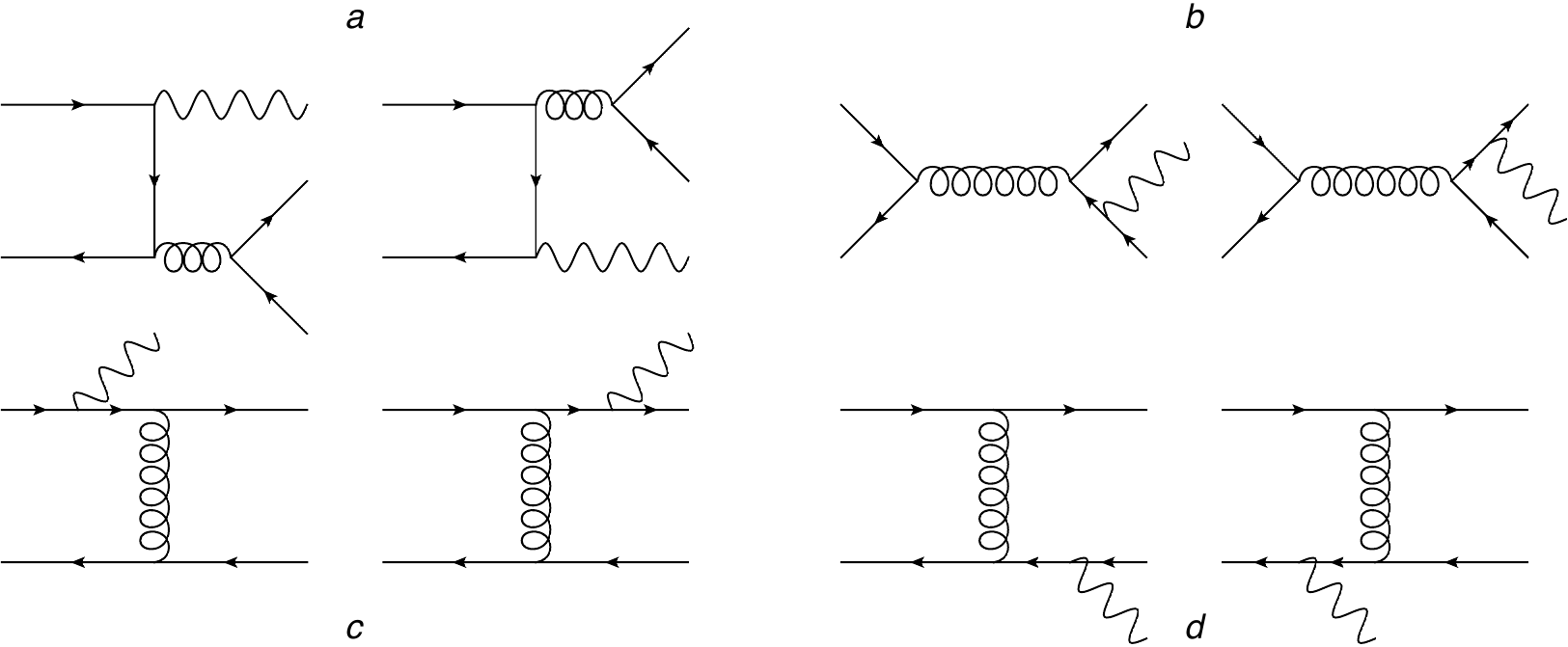}
  \caption{Diagrams which contribute to the process $q \overline{q} \rightarrow V q \overline{q}$. The $F$ functions in the cross section formula correspond to these diagrams.}
  \label{fig_Feynqqbqqb}
\end{figure}

\noindent The rest of the functions are given in the Appendix. The $(\Delta) V^{(i)}_{q\overline{q}}(s,t,u,Q^2)$ functions group all the NLO contributions proportional to $\delta(s_{23})$ arising mainly from virtual diagrams of Fig.\ref{fig_FeynVirt}, but also from infrared emission from the real ones. The $(\Delta) G_{q\overline{q}}(s,t,u,Q^2)$ are the contributions from diagrams with two final gluons of Fig.\ref{fig_FeynqqGG}, while $(\Delta) F(s,t,u,Q^2)$ functions represent the contributions from the diagrams of Fig.\ref{fig_Feynqqbqqb}, with subscripts indicating the particular groups of diagrams involved in each subprocess. Most of these functions differ by a sign between the polarized and unpolarized case, with the only exceptions being $(\Delta) F_{cc}(s,t,u,Q^2)$, $(\Delta) F_{dd}(s,t,u,Q^2)$ and $(\Delta) F^{LR}_{cd}(s,t,u,Q^2)$. This is related to helicity conservation for massless quarks. The $F^{LR}_{cd}(s,t,u,Q^2)$ function is actually equal to its polarized counterpart.

The cross section of the process $\overline{q} q \rightarrow V X$ can be obtained from Eq.(\ref{eq_masterqqb}) by substituting the corresponding couplings
\begingroup
\small
\begin{flalign}
    \overline{q} q \rightarrow V X: \quad L \leftrightarrow -R^{\dagger}. &&
\end{flalign}
\endgroup

\subsection{Quark-gluon Compton scattering}

The cross section of the Compton process is given by
\begingroup
\small
\begin{flalign}
 E_Q \frac{d (\Delta) \sigma^{qg}}{d^3 Q}=\frac{\alpha \ \alpha_s^{\overline{MS}}(\mu^2) \ C_F}{s \ (N_c^2-1)} & \Bigg\{ \delta (s_{23}) \ (\Delta) B_{qg}(s,t,u,Q^2) \sum_f \left( |L_{f1}|^2+|R_{f1}|^2 \right) \notag & \\
 +& \frac{\alpha_s^{\overline{MS}}(\mu^2)}{2 \pi} \Big\{ \Big[ \delta (s_{23}) \big[ (\Delta) V^{(3)}_{qg}(s,t,u,Q^2) + (\Delta) V^{(1)}_{qg}(s,t,u,Q^2) \sum_f (1) \big] \notag \\
 &\quad \quad + (\Delta) G_{qg}(s,t,u,Q^2) \Big] \sum_f \left( |L_{f1}|^2+|R_{f1}|^2 \right) \notag \\
 & \quad + \delta (s_{23}) (\Delta) V^{(2)}_{qg}(s,t,u,Q^2) \left(L_{11}-R_{11}\right) \sum_f \left(L_{ff}-R_{ff}\right) \Big\} \Bigg\},
\label{eq_masterqg}
\end{flalign}
\endgroup
\noindent where in this case the (polarized) unpolarized Born terms function reads 
\small
\begin{equation}
 \Delta B_{qg}(s,t,u,Q^2)=\frac{(s-t) \left(Q^2+u\right)}{s t}, \quad B_{qg}(s,t,u,Q^2)=-\frac{s^2+t^2+2 u Q^2}{s t}. 
\label{eq_Bornqg}
\end{equation}
\normalsize
\noindent The $(\Delta) G_{qg}(s,t,u,Q^2)$ terms come from diagrams obtained by crossing from those in Fig.\ref{fig_FeynqqGG}, while the $(\Delta) V^{(i)}_{qg}(s,t,u,Q^2)$ are again the $\delta(s_{23})$ contributions originating from virtual diagrams contributions, which can be obtained from Fig.\ref{fig_FeynVirt} by crossing, and from the same real contributions. Their explicit formulas are given in the Appendix. 

The $qg$ process presents the greatest difference between the polarized and unpolarized cross sections. As a matter of fact, Eq.(\ref{eq_Bornqg}) reveals that at the Born level they have two distinct behaviours: one is symmetric in the variables $s$ and $t$, while the other one is completely antisymmetric in them.

The three variations of the Compton process can be obtained from Eq.(\ref{eq_masterqg}) with the following substitutions
\begingroup
\small
\begin{flalign}
    \overline{q} g \rightarrow V X:& \quad L \leftrightarrow -R^{\dagger}, & \notag \\
    g q \rightarrow V X:& \quad t \leftrightarrow u, \quad f_1 \leftrightarrow f_2, & \notag \\
    \overline{q} g \rightarrow V X:& \quad t \leftrightarrow u, \quad f_1 \leftrightarrow f_2, \quad L \leftrightarrow -R^{\dagger}.&
\end{flalign}
\endgroup

\subsection{Gluon-gluon fusion}

The cross section of the gluon-gluon fusion process, which only contributes at NLO, is simply given by

\begingroup
\small
\begin{flalign}
 E_Q \frac{d (\Delta) \sigma^{gg}}{d^3 Q}&=\frac{\alpha \ \alpha_s^{\overline{MS}}(\mu^2) \ N_c \ C_F}{s \ (N_c^2-1)^2} \left( \frac{\alpha_s^{\overline{MS}}(\mu^2)}{2 \pi} \right) (\Delta) G_{gg}(s,t,u,Q^2) \sum_f \sum_{f'} \left( |L_{ff'}|^2+|R_{ff'}|^2 \right), &
\end{flalign}
\label{eq_mastergg}
\endgroup

\noindent where the $(\Delta) G_{gg}(s,t,u,Q^2)$ functions, corresponding to the diagrams obtained by crossing from the ones in Fig.\ref{fig_FeynqqGG}, are given in the Appendix. The $G_{gg}(s,t,u,Q^2)$ and $\Delta G_{gg}(s,t,u,Q^2)$ functions are fairly similar apart from a sign and a few additional terms. 

\subsection{Quark-quark scattering}

Finally, the last NLO contribution is the quark-quark initiated, whose cross section is given by

\begin{figure}
\centering
  \includegraphics[width=0.6\linewidth]{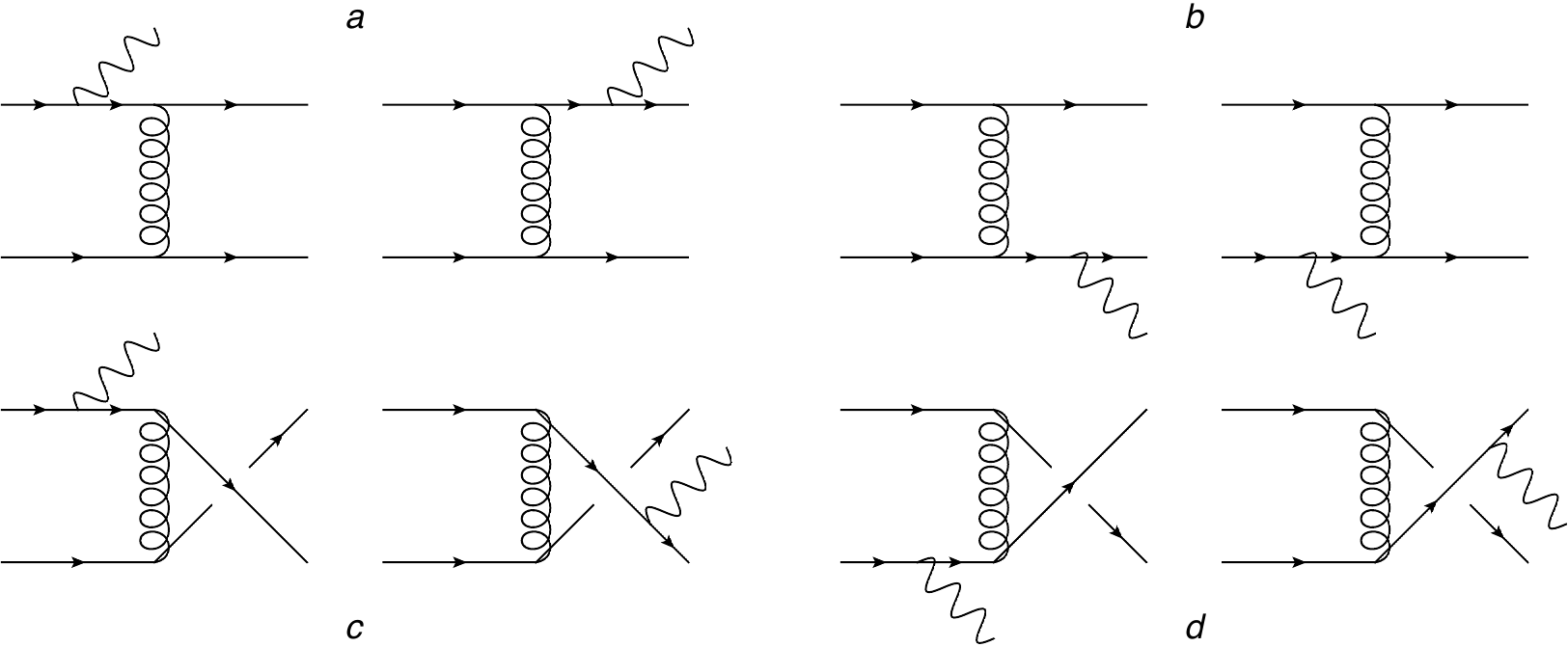}
  \caption{Diagrams corresponding to the $q q \rightarrow V q q$ process. The $H$ functions in the cross section formula correspond to these diagrams.}
  \label{fig_Feynqq}
\end{figure}

\begingroup
\small
\begin{flalign}
 E_Q \frac{d (\Delta) \sigma^{qq}}{d^3 Q}=&\frac{\alpha \ \alpha_s^{\overline{MS}}(\mu^2) \ C_F}{s \ N_c} \left( \frac{\alpha_s^{\overline{MS}}(\mu^2)}{2 \pi} \right) \left( \frac{1}{2} \right) \notag & \\
 & \times \Big\{ \big[ (\Delta) H_{aa}(s,t,u,Q^2) + (\Delta) H_{cc}(s,t,u,Q^2) \big] \sum_f \left( |L_{f1}|^2+|R_{f1}|^2 \right) \notag \\
 &+ \big[ (\Delta) H_{bb}(s,t,u,Q^2) + (\Delta) H_{dd}(s,t,u,Q^2) \big] \sum_f \left( |L_{2f}|^2+|R_{2f}|^2 \right) \notag \\
 &+ (\Delta) H_{ac}(s,t,u,Q^2) \left( |L_{21}|^2+|R_{21}|^2 \right) + (\Delta) H_{bd}(s,t,u,Q^2) \left( |L_{12}|^2+|R_{12}|^2 \right) \notag \\
 &+\Big[ (\Delta) H_{ad}(s,t,u,Q^2) + (\Delta) H_{bc}(s,t,u,Q^2) \Big] \delta_{12} \sum_f \left( |L_{f1}|^2+|R_{f1}|^2 \right) \notag \\
 &+ \Big[ (\Delta) H^{LL}_{ab}(s,t,u,Q^2) + (\Delta) H^{LL}_{cd}(s,t,u,Q^2) \Big] \left( L_{11} L_{22}+R_{11} R_{22} \right) \notag \\
 &+ \Big[ (\Delta) H^{LR}_{ab}(s,t,u,Q^2) + (\Delta) H^{LR}_{cd}(s,t,u,Q^2) \Big] \left( L_{11} R_{22}+R_{11} L_{22} \right) \Big\},
\label{eq_masterqq}
\end{flalign}
\endgroup

\noindent where the $(\Delta) H(s,t,u,Q^2)$ functions correspond to the contributions from the diagrams in Fig.\ref{fig_Feynqq}. The global $1/2$ factor is a statistical factor needed to avoid double counting events due to the full phase space integration and quantum number summation over the two quarks in the final state. Most of the $H$ functions are related to the $F$ ones from the quark-antiquark process. The $H_{ac}(s,t,u,Q^2)$, $H_{bd}(s,t,u,Q^2)$, $H_{ad}(s,t,u,Q^2)$, $H_{bc}(s,t,u,Q^2)$, $H^{LR}_{ab}(s,t,u,Q^2)$ and $H^{LR}_{cd}(s,t,u,Q^2)$ functions are equal to their polarized counterparts, while the $H^{LL}_{ab}(s,t,u,Q^2)$ and $H^{LL}_{cd}(s,t,u,Q^2)$ differ only by a sign. The $H_{bb}(s,t,u,Q^2)$ and $H_{dd}(s,t,u,Q^2)$ functions, as it is the case with $F_{cc}(s,t,u,Q^2)$ and $F_{dd}(s,t,u,Q^2)$, have significant differences with their polarized counterparts.

The cross section of antiquark-antiquark annihilation is obtained from Eq.(\ref{eq_masterqq}) by the substitution
\begingroup
\small
\begin{flalign}
    \overline{q} \overline{q} \rightarrow V X: \quad L \leftrightarrow -R^{\dagger}. &&
\end{flalign}
\endgroup

With the $q \overline{q}$ and $q q$ cross sections it is possible to calculate the non-singlet cross section, which is given by
\small
\begin{equation}
 E_Q \frac{d (\Delta) \sigma^{NS}}{d^3 Q}=E_Q \frac{d (\Delta) \sigma^{q \overline{q}}}{d^3 Q}-E_Q \frac{d (\Delta) \sigma^{q q}}{d^3 Q}.
\end{equation}
\normalsize

The non-singlet polarized cross section for virtual photon production has been calculated by Field et al. in Ref.\cite{Chang:1997ik}. However, our results do not fully agree with theirs because they assumed that the $\Delta H^{LL}_{ab}(s,t,u,Q^2)$, $\Delta H^{LL}_{cd}(s,t,u,Q^2)$, $\Delta H^{LR}_{ab}(s,t,u,Q^2)$ and $\Delta H^{LR}_{cd}(s,t,u,Q^2)$ contributions of the $q q$ process cancel against the corresponding $\Delta F^{LL}_{cd}(s,t,u,Q^2)$ and $\Delta F^{LR}_{cd}(s,t,u,Q^2)$ functions of the $q \overline{q}$ process, which is not correct since they actually differ by a sign and add up to an additional contribution. This error was carried over from a mistake in the original unpolarized EMP calculation \cite{Ellis:1981hk}, which it is pointed out in Ref.\cite{Gonsalves:1989ar}.

\section{Phenomenological Results}
\label{sect3}

In this section we compute the corresponding cross sections to evaluate the impact of the polarized NLO corrections. We apply them to  RHIC kinematics, that is, polarized proton-proton collisions at a center-of-mass energy of $\sqrt{S}=510$ GeV. We use the MMHT2014nlo \cite{Harland-Lang:2014zoa} and DSSV \cite{deFlorian:2009vb} PDFs sets for the unpolarized and polarized processes, respectively. For the weak gauge boson we use the masses $M_W=80.379$ GeV and $M_Z=91.1876$ GeV, with an electromagnetic coupling constant $\alpha=1/127.918$ and the Weinberg angle given by $\sin{\theta_W}^2=0.2312$. The CKM matrix is constructed from the values $|V_{ud}|=0.9742$, $|V_{us}|=0.2243$, $|V_{cd}|=0.218$ and $|V_{cs}|=0.996$ using the universality property. The number of active quark flavors are limited by the thresholds set by their mass, namely $m_c=1.275$ GeV and $m_b=4.18$ GeV. The QCD coupling $\alpha_s$ is evaluated at NLO with the same quark thresholds, with a value of $\alpha_s(M_Z)=0.127$.

\begin{figure}[h]
\centering
  \includegraphics[width=0.5\linewidth]{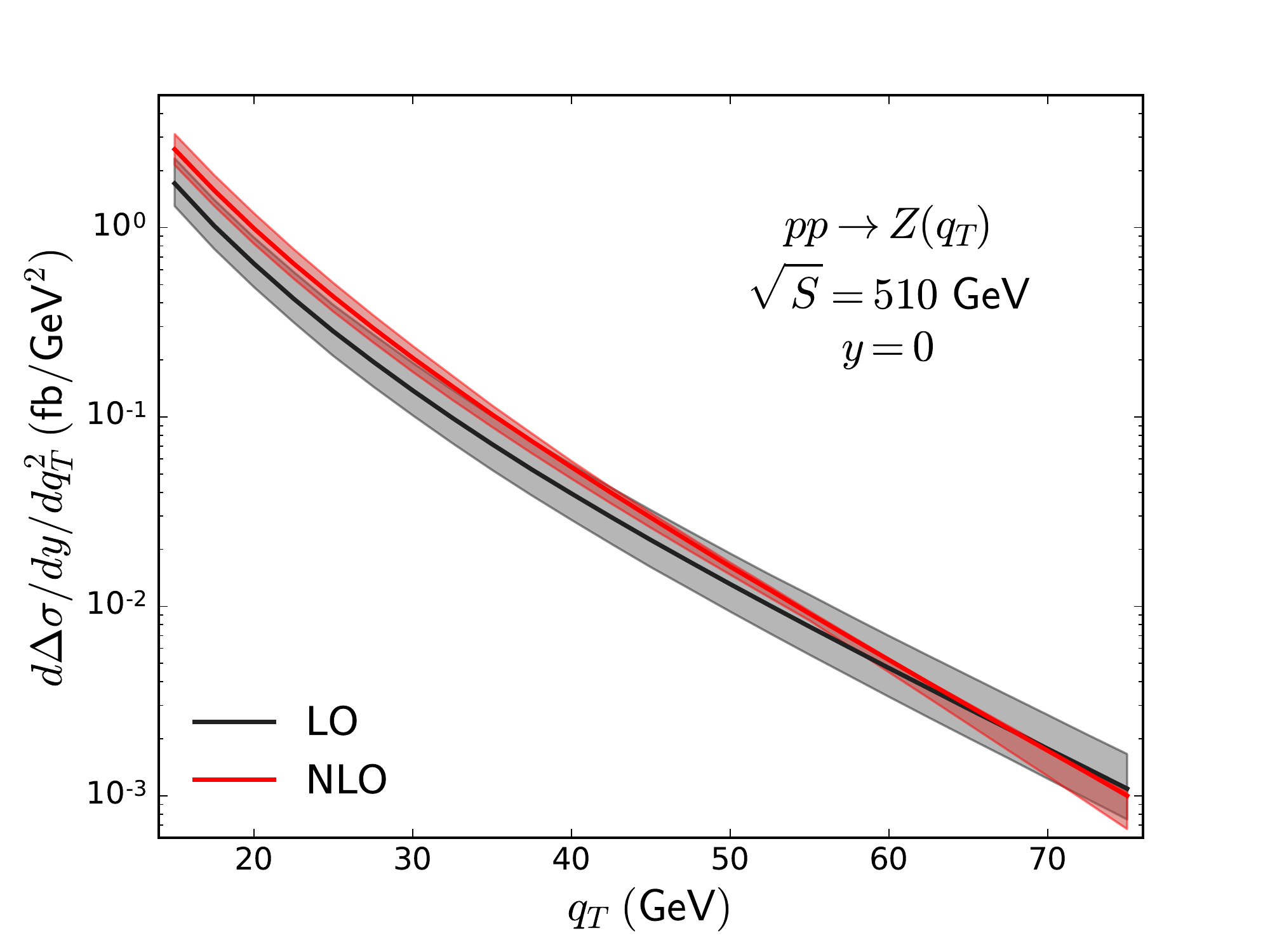}
  \caption{On-shell $Z$ production cross section in $pp$ collisions at $510$ GeV. Both LO and NLO are presented with their respective confidence bands.}
  \label{fig_plotZunc}
\end{figure}

We begin by analysing the reduction of theoretical uncertainties. These are estimated by studying the dependence of the complete NLO results on the renormalization and factorization scales due to the truncation of the perturbative series expansion. In Fig.\ref{fig_plotZunc} we present the polarized $q_T$ distributions of on-shell $Z$ boson production at both LO and NLO accuracy at $y=0$, with their respective confidence bands. Since in this process we have two physical scales given by $Q$ and $q_T$, it is convenient to define a physical scale in between them given by $Q_s^2=(Q^2+q_T^2)/4$. The central curves are obtained by fixing the renormalization and factorization scales to $\mu_R=\mu_F=Q_s$. The bands are obtained by varying $\mu_R$ and $\mu_F$ simultaneously and independently in the range $0.5 Q_s \ \leq \ \mu_R,\mu_F \ \leq \ 2 Q_s$ with the constraint $0.5 \leq \mu_F/\mu_R \leq 2$. As expected in perturbative QCD, the second order corrections yield a significant contribution to the Born cross section while reducing uncertainties considerabily. 

\begin{figure}[!b]
\centering
  \includegraphics[width=0.5\linewidth]{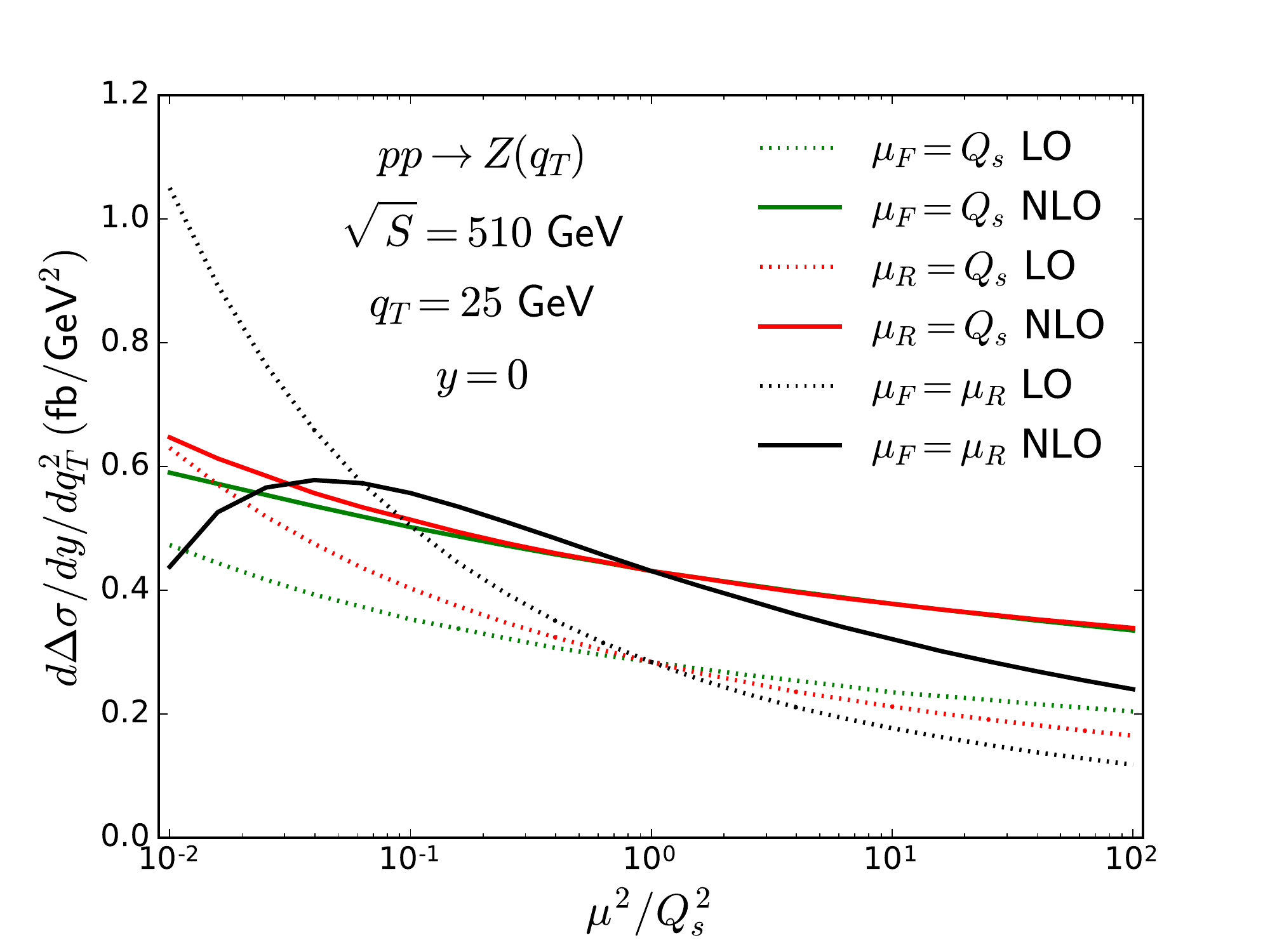}
  \caption{Cross section dependence on the renormalization and factorization scales. The green curves are for $\mu=\mu_R$ and $\mu_F=Q_s$, the red ones for $\mu=\mu_F$ and $\mu_R=Q_s$, and the black ones for $\mu=\mu_F=\mu_R$.}
  \label{fig_plotZscldep}
\end{figure}

For a more detailed analysis on the impact of the two scales, in Fig.\ref{fig_plotZscldep} we study the scale dependence of the cross section at a fixed value of $q_T$. The LO curves show stronger dependence on the scale. In this case the $\mu_R$ dependence corresponds simply to the variations of $\alpha_s$ with the scale, while $\mu_F$ dependence arises exclusively from the PDFs. At NLO these dependences are corrected by logarithmic terms proportional to the $\beta$ function and the splitting functions, respectively, yielding more stable results. We do not show the results for other particular values of $q_T$ since they are fairly similar. Throughout the rest of this section we will set the two scales equal to $Q_s$.

The size of the higher order QCD corrections to hadronic processes are usually presented in terms of the `$K$-factor', the ratio between the NLO results over the LO ones. However, in order to make this ratio meaningful in the polarized case, the same NLO-evolved parton densities are used to calculate both NLO and LO quantities \cite{deFlorian:1998qp}. This is due to the low constraints to polarized PDFs available from experimental data, especially to $\Delta g$, which can in some cases give very different results when fits are performed at LO or at NLO. For this reason, polarized $K$-factors can get artificially large or small when the gluons are involved.

\begin{figure}
\begin{subfigure}{0.49\textwidth}
\includegraphics[width=\linewidth]{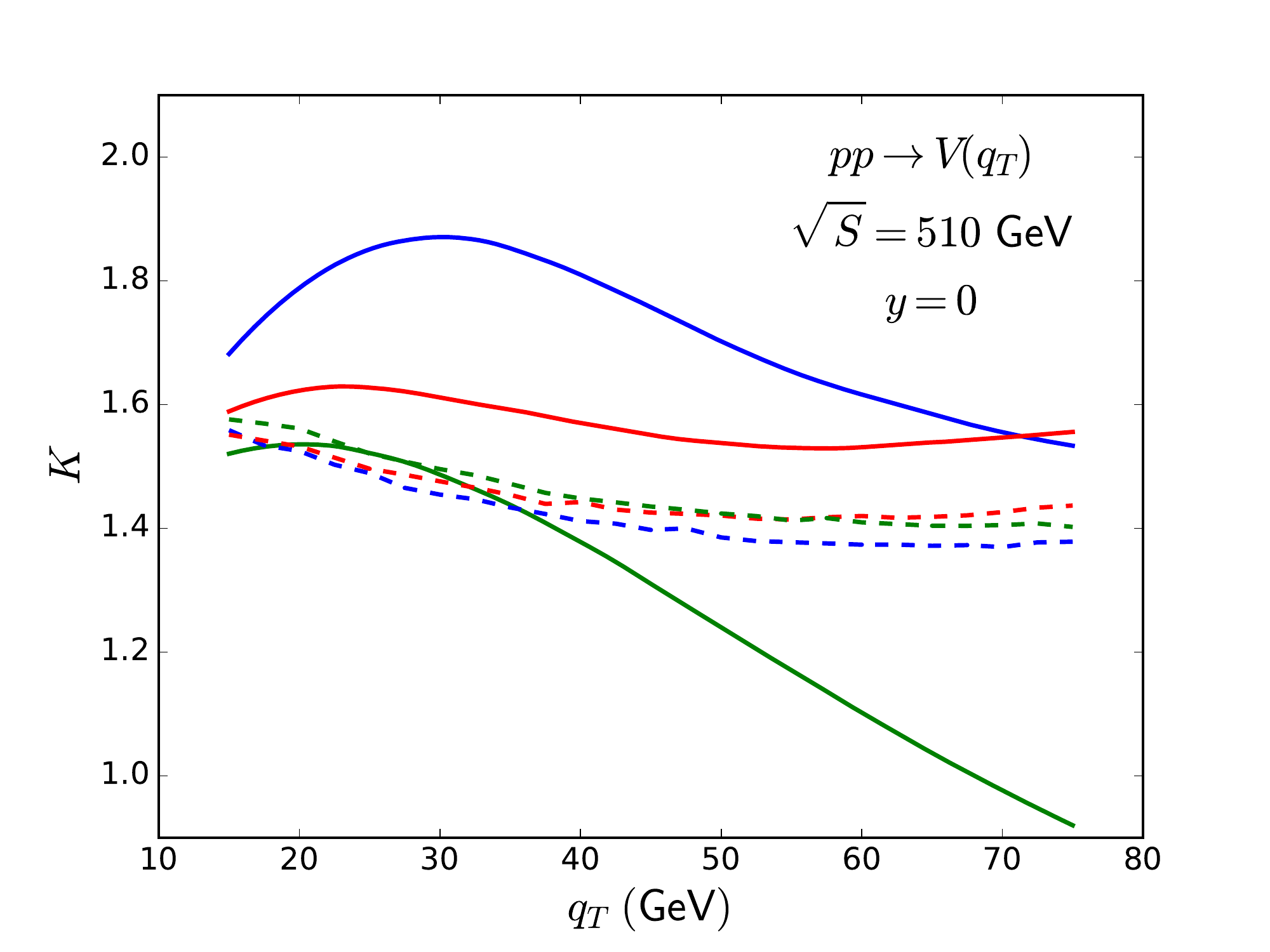}
\end{subfigure}
\hspace*{\fill} 
\begin{subfigure}{0.49\textwidth}
\includegraphics[width=\linewidth]{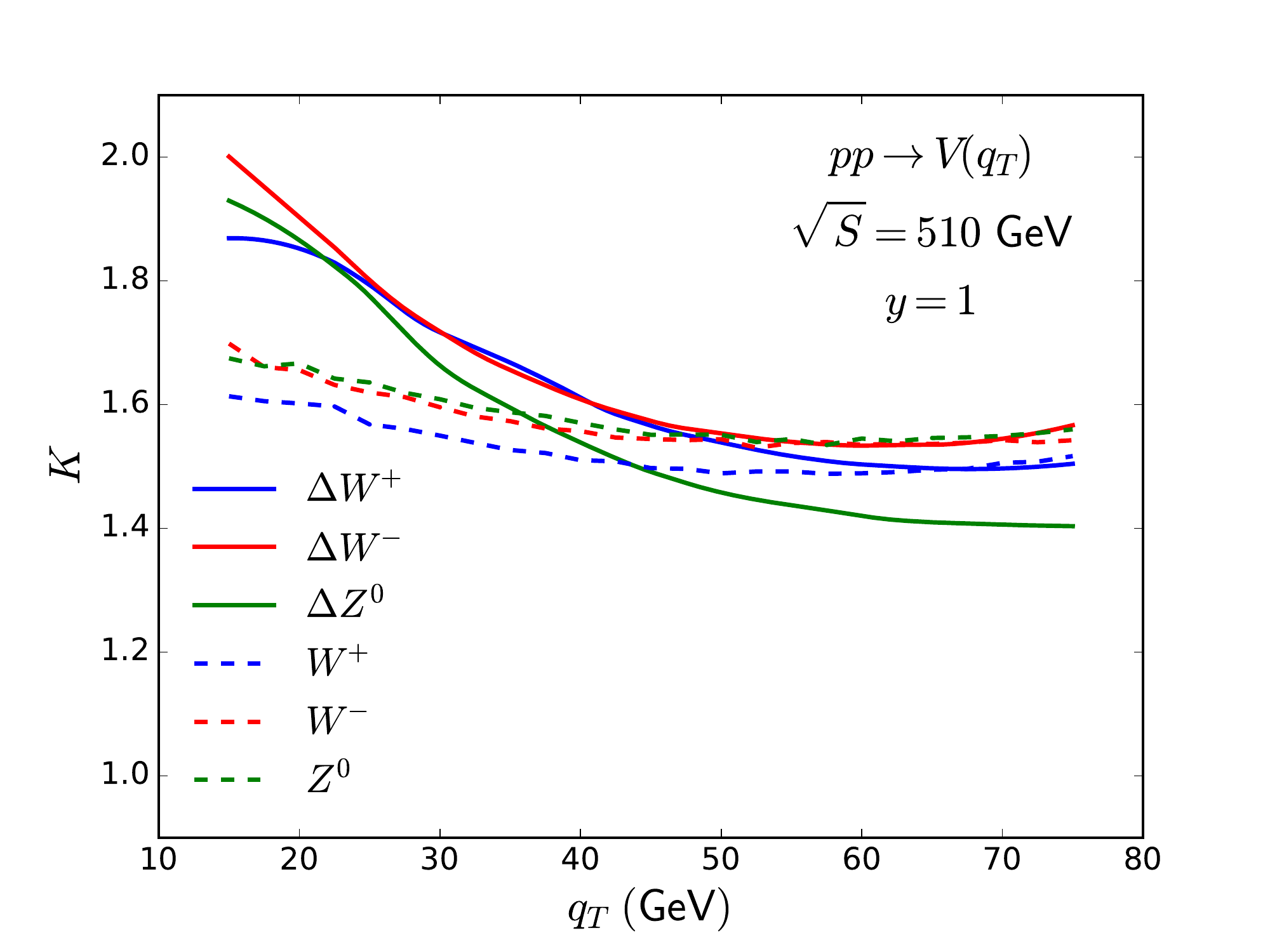}
\end{subfigure}
\caption{NLO $K$ factors for on-shell weak boson $V$ production with rapidities $y=0$ and $y=1$. The solid and dashed curves correspond to the polarized and unpolarized processes, respectively.} \label{fig_plotkf}
\end{figure}

In Fig.\ref{fig_plotkf} we present the $K$-factors for the production of on-shell weak bosons at fixed rapidity $y=0$ and $y=1$, for both the polarized and unpolarized cases. The polarized $K$ factors reveal a larger contribution of the NLO corrections for $W$ production, and they show a stronger variation along $q_T$ compared to their unpolarized counterparts. The polarized $Z$ production present a remarkably lower $K$ factor at larger $q_T$. At rapidity $y=1$ the $K$-factors are slightly higher, specially at low $q_T$. In this case the bosons $W^{+}$ and $W^{-}$ show similar behaviours.

\begin{figure}[h]
\begin{subfigure}{0.49\textwidth}
\includegraphics[width=\linewidth]{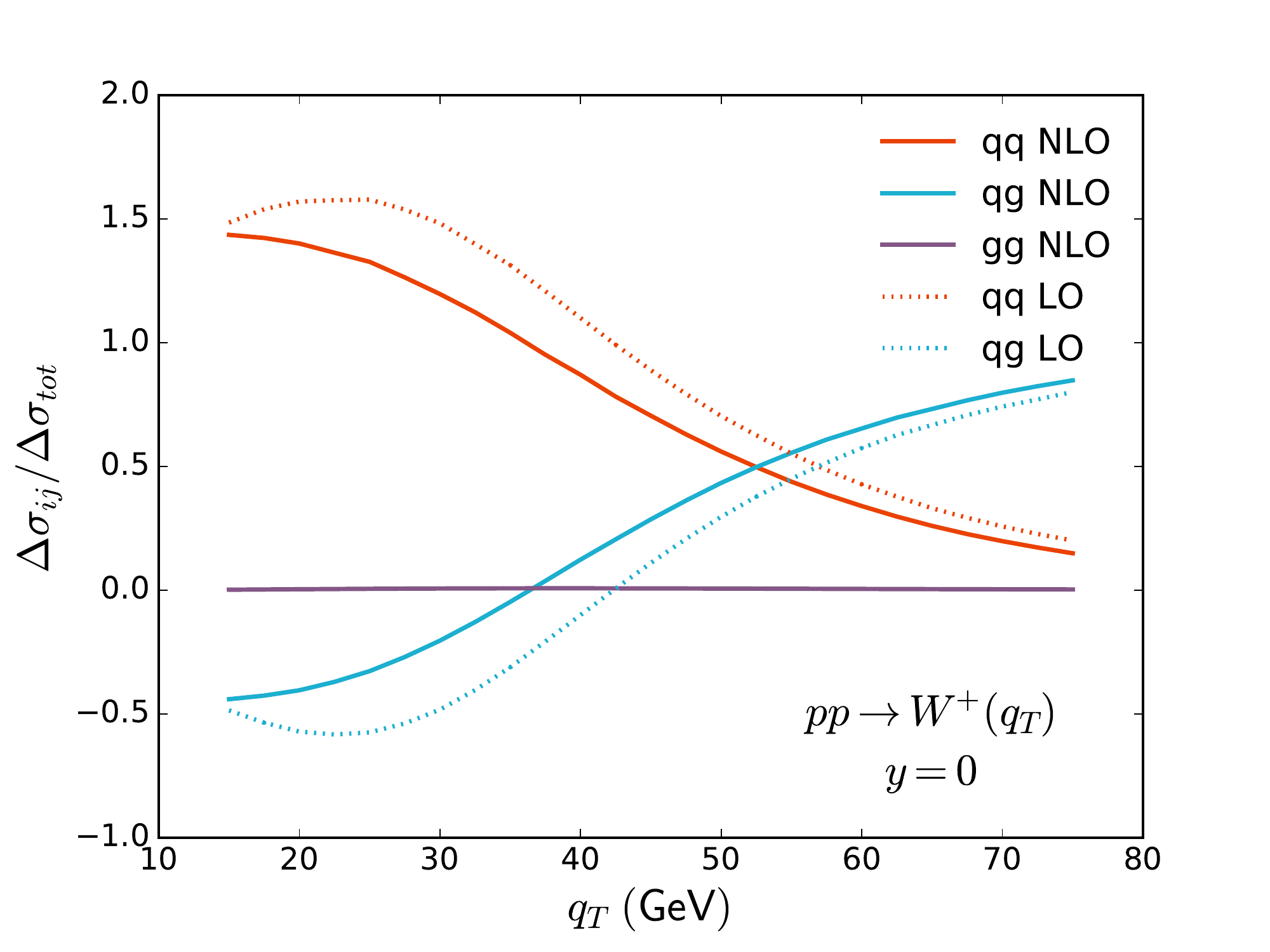}
\end{subfigure}
\hspace*{\fill} 
\begin{subfigure}{0.49\textwidth}
\includegraphics[width=\linewidth]{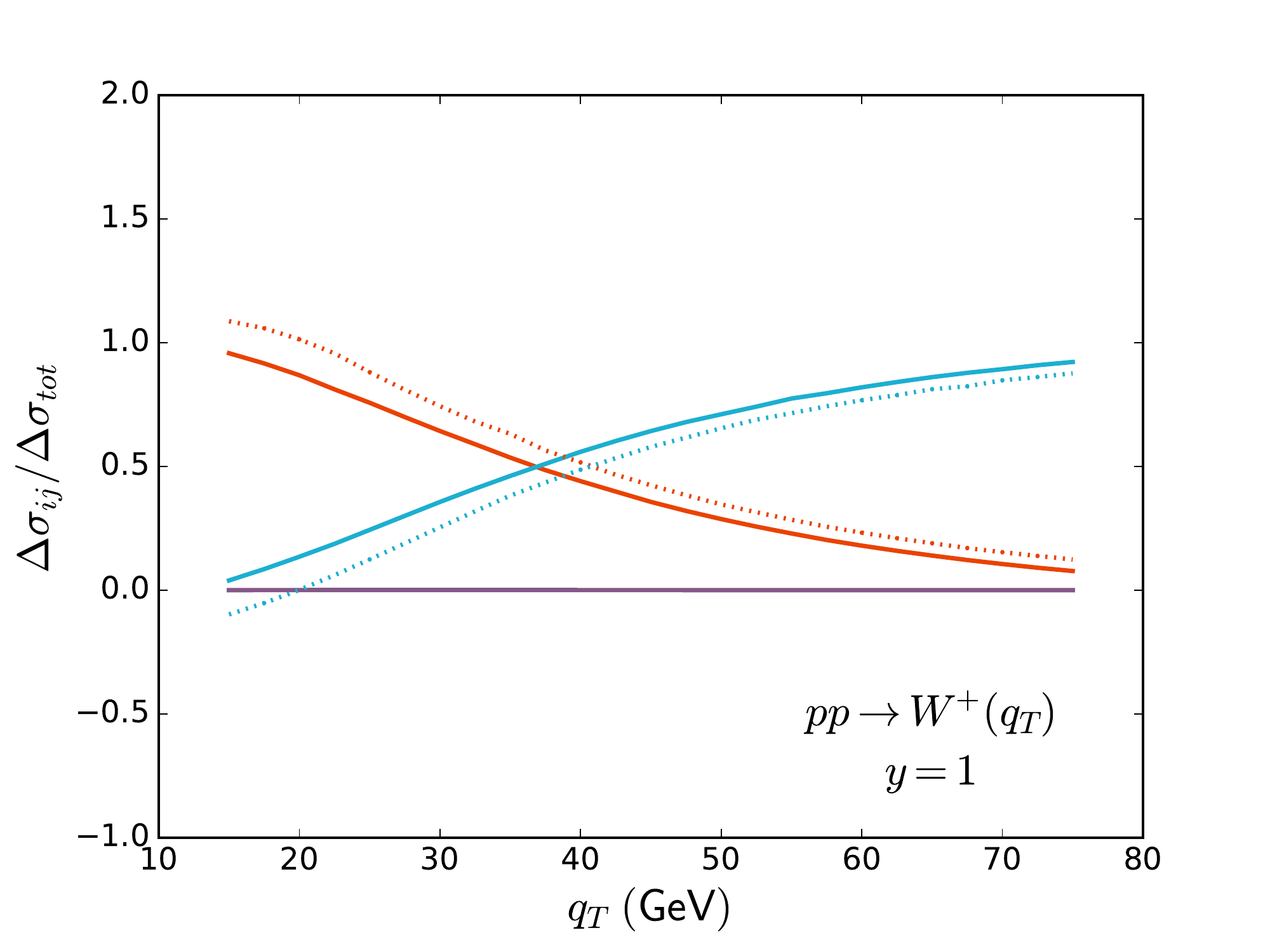}
\end{subfigure}
\caption{Sub-channel contributions to on-shell $W^{+}$ boson production with rapidities $y=0$ and $y=1$. Both NLO and LO decompositions are given.} \label{fig_plotchann}
\end{figure}

The $q_T$ dependence of the polarized $K$ factors is related to the different behaviour of the particular sub-channels contributing to the process. For instance, the changes in the concavity in the curves for $y=0$ (barely visible in the $Z$ boson case) are related to the sign change  of the $qg$-channel contribution. This can be observed in Fig.\ref{fig_plotchann} for $W^{+}$ production, where we present the sub-channel polarized cross section ratios. In all cases the cross section is governed by the quark initiated channels at low $q_T$, but the $qg$-channel becomes the most relevant one at high $q_T$. The $q_T$ range at which this dominance begins depends on the particular boson studied and its rapidity, taking place at lower $q_T$ values in the case $y=1$. The $gg$ contribution, that only arises at order $\mathcal{O}(\alpha_s^2)$, is always negligible at the energy of this process. 

\begin{figure}[h]
\begin{subfigure}{0.49\textwidth}
\includegraphics[width=\linewidth]{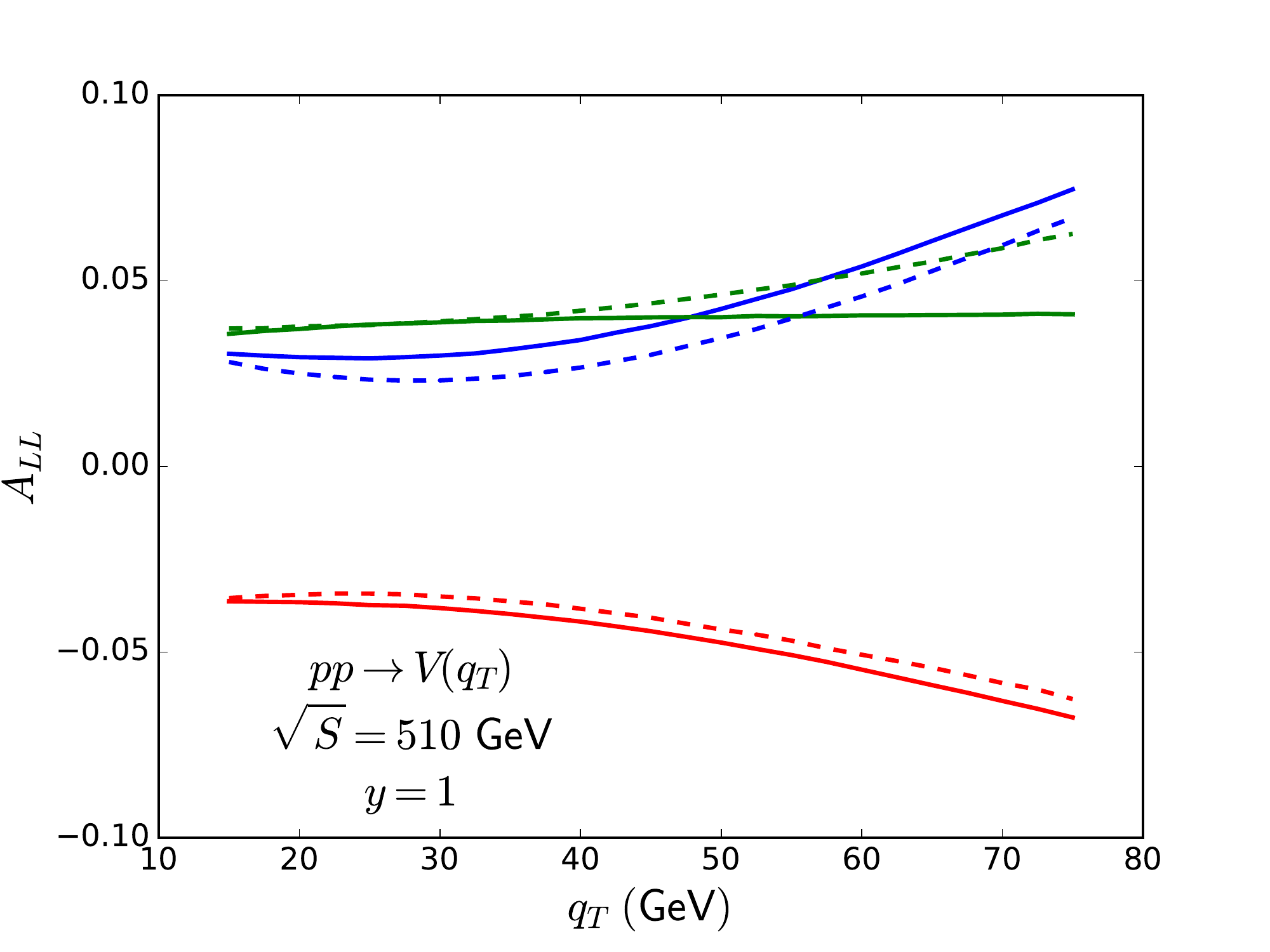}
\end{subfigure}
\hspace*{\fill} 
\begin{subfigure}{0.49\textwidth}
\includegraphics[width=\linewidth]{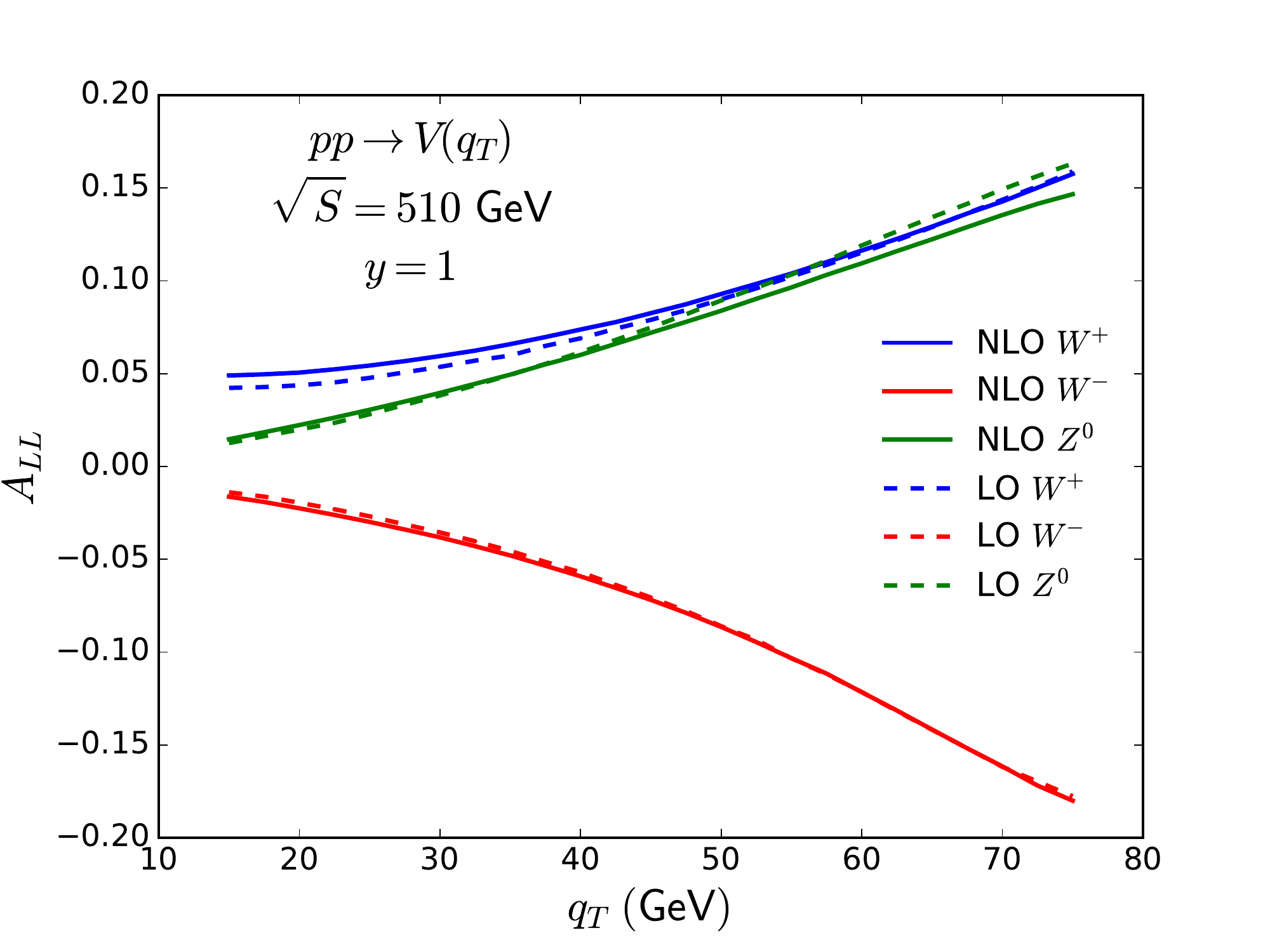}
\end{subfigure}
\caption{Longitudinal double spin asymmetries in on-shell weak boson $V$ production at rapidities $y=0$ and $y=1$.} \label{fig_plotall}
\end{figure}

Finally in Fig.\ref{fig_plotall} we show the longitudinal double spin asymmetries for weak boson production at different rapidities, which is defined as

\begin{equation}
 A_{LL}=\frac{d \Delta \sigma}{d \sigma}.
\end{equation}

\noindent The asymmetries grow with $q_T$, reaching sizeable values, and are larger at higher rapidity. The NLO corrections to the asymmetries are generally low, especially at rapidity $y=1$. A similar behaviour has been observed in many other processes, where QCD corrections tend to compensate in the ratios of polarized and unpolarized cross sections. However, in the case of $Z$ production the NLO corrections in the polarized case are clearly bigger than the unpolarized ones, accounting for lower values of asymmetry at large $q_T$. This can already be seen in Fig.\ref{fig_plotkf} since the ratios of the asymmetries at the two orders are equal to the ratio of the respective $K$ factors, following the relation $A^{NLO}=A^{LO} \ \Delta K/K$, and the K factors of $Z$ production are the ones that show larger discrepancies between the polarized and unpolarized cases, especially at $y=0$.

\section{Summary}
\label{sec:summa}

In this work, we have presented the first complete calculation at next-to-leading order in perturbative QCD of the cross section for the production of electroweak gauge bosons with large transverse momentum $q_T$ in double polarized hadronic collisions. The calculation was done fully in the consistent HVBM scheme for the treatment of $\gamma^5$ and the tensor $\epsilon_{\mu \nu \rho \sigma}$ in dimensional regularization, carefully dealing with issues arising from the $(d-4)$-dimensional part of the momenta and $\gamma^{\mu}$ matrices. We also kept the unpolarized results, easily obtained as a by-product of the polarized calculation, corroborating the previous well-known results \cite{Gonsalves:1989ar}, but this time fully within the HVBM scheme. The polarized results are also in agreement with a previous partial calculation of the non-singlet polarized production of virtual photons \cite{Chang:1997ik}, aside from a missed contribution in their work.

Using our results, we studied in some detail the phenomenological impact of the polarized NLO corrections in weak boson production at RHIC, that is, polarized $pp$ collisions with a center-of-mass energy of 510 GeV. We analysed the scale dependence in polarized $Z$ boson production to assess the reliability of the theoretical predictions. We found that the NLO correction have a greatly reduced scale dependence with respect to the Born results. We also studied the polarized $K$ factors, the ratio between the NLO and LO results, comparing them to their unpolarized counterparts. The polarized $K$ factors show larger variations and a more complex dependence in $q_T$, having a slightly higher value than the unpolarized ones except for the $Z$ boson case. The effect of NLO corrections on the double spin asymmetries is generally low for $W$ production, but visibly affects the $Z$ production at higher $q_T$. 

\section*{Acknowledgements}
\label{sec:akno}

We are grateful to Werner Vogelsang for helpful discussions. This work has been partially supported by Conicet and ANPCyT.

\appendix
\section{Appendix}

In this appendix we present the formulae for the functions associated with the NLO contributions in the cross sections of Eqs.(\ref{eq_masterqqb}), (\ref{eq_masterqg}), (\ref{eq_mastergg}), and (\ref{eq_masterqq}). They are expressed in terms of the invariants $s, t, u, Q^2$ and $s_{23}$, but we also define some frequently occurring transcendental functions and denominators. The denominators are:
\small
\begin{equation}
    \begin{aligned}[c]
        \text{ds} &= s+Q^2-s_{23},\\
        \text{dq} &= s_{23} Q^2-t u, \\
	\lambda &= \sqrt{(t+u)^2-4 s_{23} Q^2},
    \end{aligned}
\qquad
    \begin{aligned}[c]
        \text{dt} &= s_{23}-t,\\
        \text{du} &= s_{23}-t,
    \end{aligned}
\qquad
    \begin{aligned}[c]
        \text{dst} &= s-s_{23}+t,\\
        \text{dsu} &= s-s_{23}+u.
    \end{aligned}
\end{equation}
\normalsize

The first group of transcendental functions involves some recurring logarithms:  
\small
\begin{equation}
    \begin{aligned}[c]
        L_s &= \ln \left[ \frac{s}{Q^2}\right],\\
        L_t &= \ln \left[ \frac{t}{Q^2}\right],\\
        L_u &= \ln \left[ \frac{u}{Q^2}\right],\\
        L_{s_{23}} &= \ln \left[\frac{s_{23}}{Q^2}\right],\\
        L_A &= \ln \left[ \frac{A}{Q^2}\right],
    \end{aligned}
\qquad
    \begin{aligned}[c]
        L_{\lambda t} &= \ln \left[\frac{s \ Q^2 \left( s_{23}-t\right)^2}{\left[ s_{23}(2 \  Q^2-t)-Q^2 \ t\right]^2}\right],\\
        L_{\lambda u} &= \ln \left[\frac{s \ Q^2 \left( s_{23}-u\right)^2}{\left[ s_{23}(2 \  Q^2-u)-Q^2 \ u\right]^2}\right], \\
        L_{tu} &= \ln \left[\frac{tu-s_{23} \ Q^2}{\left( s_{23}-t\right) \left( s_{23}-u\right)}\right], \\
        L_{stu} &= \ln \left[\frac{s \ Q^2}{\left( s_{23}-t\right) \left( s_{23}-u\right)}\right],
    \end{aligned}
\qquad
    \begin{aligned}[c]
        L_{st} &= \ln \left[\frac{s \ t^2}{Q^2\left( s_{23}-t\right)^2}\right],\\
        L_{su} &= \ln \left[\frac{s \ u^2}{Q^2\left( s_{23}-u\right)^2}\right], \\
	L_{\lambda} &=  \ln \left[ \frac{s+Q^2-s_{23}+\lambda}{s+Q^2-s_{23}-\lambda} \right], \\
	L_{R} &= \ln \left[ \frac{\mu_R^2}{Q^2}\right],\\
	L_F &= \ln \left[ \frac{\mu_F^2}{Q^2}\right] \, ,
    \end{aligned}
\end{equation}
\normalsize
with $\mu_R$ and $\mu_F$ the renormalization and factorization scales, respectively.
The rest of the transcendental functions include dilogarithms that appear in the virtual contributions:
\small
\begin{equation}
    \begin{aligned}[c]
        L_{1t} &= \operatorname{Li}_2 \left[ \frac{Q^2}{Q^2-t}\right]+\frac{1}{2} \ln^2 \left[ \frac{Q^2}{Q^2-t}\right],\\
        L_{1u} &= \operatorname{Li}_2 \left[ \frac{Q^2}{Q^2-u}\right]+\frac{1}{2} \ln^2 \left[ \frac{Q^2}{Q^2-u}\right],
    \end{aligned}
\qquad
    \begin{aligned}[c]
        L_{2t} &= \operatorname{Li}_2 \left[ \frac{Q^2}{s}\right]+\frac{1}{2} L_s^2+L_s \ln^2 \left[ \frac{-t}{s-Q^2}\right],\\
	L_{2u} &= \operatorname{Li}_2 \left[ \frac{Q^2}{s}\right]+\frac{1}{2} L_s^2+L_s \ln^2 \left[ \frac{-u}{s-Q^2}\right].
    \end{aligned}
\end{equation}
\normalsize

Finally, some expressions include `+' distributions in the variable $s_{23}$. These terms originate from the soft singularities appearing as poles in $s_{23}$. Within dimensional regularization, these poles are explicitly made manifest with the identity 
\begingroup
\small
\begin{flalign}
	\left( \frac{1}{s_{23}} \right)^{1+\epsilon}=-\frac{1}{\epsilon} \ \delta\left(s_{23}\right) \left[ 1-\epsilon \ \ln A+\frac{1}{2} \epsilon^2 \ln A^2\right]+\left(\frac{1}{s_{23}}\right)_{A+}-\epsilon \left(\frac{\ln s_{23}}{s_{23}}\right)_{A+}+\mathcal{O}(\epsilon^2),
\end{flalign}
\endgroup

\noindent where $A$ is the upper integration limit in Eq.(\ref{eq_ints23}). The `+' distributions yield finite contributions and are integrated with the following rules:
\begingroup
\small
\begin{flalign}
	&\int^A_0 ds_{23} \ \Big(\frac{1}{s_{23}} \Big)_{A+} f(s_{23}) = \int^A_0 ds_{23} \ \frac{[f(s_{23})-f(0)]}{s_{23}}, & \notag \\
 	&\int^A_0 ds_{23} \ \Big(\frac{\ln s_{23}}{s_{23}} \Big)_{A+} f(s_{23}) = \int^A_0 ds_{23} \ \frac{[f(s_{23})-f(0)]}{s_{23}} \ \ln(s_{23}), &
\end{flalign}
\endgroup

\noindent where $f(s_{23})$ is any function that depends on $s_{23}$.

\subsection{Quark-antiquark annihilation and scattering}

In this subsection we present the functions corresponding to the contributions to the quark-antiquark annihilation and scattering processes in both the polarized and unpolarized cases.  

\subsubsection{Polarized $q \overline{q} \rightarrow V X$}

We first present the $\delta(s_{23})$ contributions arising from virtual and real contributions
\begingroup
\small
\begin{flalign}
 \Delta V^{(1)}_{q\overline{q}}(s,t,u,Q^2)=\Delta B_{q\overline{q}}(s,t,u,Q^2) \left[ -\frac{5}{9}+\frac{L_{A}-L_{R}}{3} \right], &&
\end{flalign}
\endgroup
\begingroup
\small
\begin{flalign}
 \Delta V^{(2)}_{q\overline{q}}(s,t,u,Q^2)=\frac{(s+Q^2)}{(s-Q^2)} \left[ 1-\frac{L_{s} Q^2}{(s-Q^2)} \right], &&
\end{flalign}
\endgroup
\begingroup
\small
\begin{flalign}
 \Delta V^{(3)}_{q\overline{q}}(s,t,u,Q^2)=&\Delta B_{q\overline{q}}(s,t,u,Q^2) \Bigg\{ C_F \bigg[ (2 L_{A}+L_{t}+L_{u})^2 -4 L_{A} \left(L_F-L_{s}+2 L_{t}+2 L_{u}\right) \notag & \\
 &\quad- L_F \left(3-2 L_{t}-2 L_{u} \right)-2 L_s \left(L_t+L_u \right)-8+\pi ^2 \bigg] - C_A \bigg[(L_{u}+L_{t}+L_{A})^2 \notag \\
 &\quad+L_{A} \left(\frac{11}{6}+2 L_{s}-4 L_{t}-4 L_{u}\right)-\frac{11}{6} L_{R} \notag \\
 &\quad -L_{1t}-L_{1u}-L_s \left(L_{t}+L_{u} \right)-\frac{67}{18}+\frac{\pi^2}{3} \bigg] \Bigg\} \notag \\
 &-\frac{L_u}{(s+t)} \left[C_F \left( 3 s+u\right)+\frac{C_F Q^2 s}{(s+t)}+C_A \ u \right]+\frac{C_F \ s^2}{(s+t) t} \notag \\
 &-\frac{L_t}{(s+u)} \left[C_F \left( 3 s+t\right)+\frac{C_F Q^2 s}{(s+u)}+C_A \ t \right]+\frac{C_F \ s^2}{(s+u) u} \notag \\
 &-\left( 2 C_F-C_A \right) \left[ \frac{(s+u)^2+s^2}{t u} \left( L_{1t}-L_{2t} \right)+\frac{(s+t)^2+s^2}{t u} \left( L_{1u}-L_{2u} \right)\right] \notag \\
 &-\frac{2 \left(2 C_F-C_A \right) s}{(t+u)} \left[ 1+L_s+\frac{L_s Q^2}{(t+u)}\right] + \left(C_F -C_A \right) \left( \frac{t}{u}+\frac{u}{t} \right)-C_A  s \left( \frac{1}{t}+\frac{1}{u} \right).
\end{flalign}
\endgroup

\noindent Here the $\Delta B_{q\overline{q}}(s,t,u,Q^2)$ is the function representing the Born contribution defined in Eq.(\ref{eq_Bornqqb}). The function $\Delta V^{(2)}_{q\overline{q}}(s,t,u,Q^2)$ arises from the triangle quark loops diagrams, which only contributes to $Z$ production.

The following function is originated from the quark-antiquark annihilation $q \overline{q} \rightarrow V g g$ diagrams in Fig.\ref{fig_FeynqqGG}:

\begingroup
\small
\allowdisplaybreaks
\begin{flalign}
 \Delta G_{q\overline{q}}(s,t,u,Q^2)=&-\frac{t^2+u^2+2 s \left(t+u+s \right)}{2 t u} \bigg\{ \left(8 C_F-2 C_A\right) \left( \frac{L_{s_{23}}}{s_{23}}\right)_{A+} \notag & \\
 &\quad-\left( \frac{1}{s_{23}}\right)_{A+} \left[2 C_F \left(2 L_F-L_{tu}\right)-\left(2 C_F-C_A\right) \left(2 L_{stu}-L_{tu}\right)+\frac{11}{6} C_A\right] \bigg\} \notag \\
 &+\frac{C_F}{\text{dq}} \left[ \frac{2 Q^2 u}{t}+s-Q^2 \right]-\frac{C_F \left(L_F-L_{s_{23}}-L_{tu}\right)}{\text{dq}} \Bigg[ \frac{4 s^2 \left(s-s_{23}\right)}{t u} \notag \\
 &\quad +\frac{4 s \left(4 s-2 s_{23}+3 u\right)-4 u \left(s_{23}-u\right)}{t}+4 s+s_{23} \Bigg] \notag \\
 &-\frac{s \left(2 C_F u-C_A s\right)}{2 \text{dt}^2 u}-\frac{C_F \left(L_F-L_{s_{23}}\right) s}{\text{dt}^2} \notag \\
 &-\frac{\left(2 C_F-C_A\right) \left(L_{s_{23}}+L_{stu}-L_{tu}\right)}{\text{dt}} \left[\frac{s^2}{\text{du} \ t}-\frac{(s+u)^2}{t u}+1\right] \notag \\
 &+\frac{C_F \left(L_F-L_{s_{23}}\right) \left(s+s_{23}+Q^2\right)}{\text{dt} \ t}-\frac{2 C_F}{\text{dt}} \left( \frac{u+s}{t}-\frac{s}{u} \right)-\frac{C_A s}{2 \text{dt}} \left( \frac{3 s}{t u}+\frac{4}{u} \right) \notag \\
 &-\frac{\left[3 C_F \left(L_F-L_{s_{23}}\right)+C_A\right] \left(Q^2-t\right)}{3 t^2}+\frac{C_F \left(s-2 s L_{tu}+s_{23} L_{tu}\right)}{t u} \notag \\
 &-\frac{3 C_F \left(2+L_F-L_{s_{23}}+4 L_{tu}\right)+C_A}{3 t} + \langle t \Leftrightarrow u \rangle.
\end{flalign}
\endgroup

\noindent In this function we have implicitly included a $1/2$ factor to avoid double counting due to the summation over all the final states of the two indistinguishable gluons, and the $\langle t \Leftrightarrow u \rangle$ indicates that the whole function needs to be repeated switching the variables $t$ and $u$. The annihilation process also contributes some of the $\delta(s_{23})$ terms which are already included in $\Delta V^{(3)}_{q\overline{q}}(s,t,u,Q^2)$.

We now proceed to present the functions corresponding to the $q \overline{q} \rightarrow V q \overline{q}$ process diagrams represented in Fig.\ref{fig_Feynqqbqqb}:

\begingroup
\small
\begin{flalign}
 \Delta F_{aa}(s,t,u,Q^2)=\frac{1}{3} \left[-\left( \frac{1}{s_{23}}\right)_{A+} \frac{t^2+u^2+2 s (s+t+u)}{t u}+Q^2 \left(\frac{1}{t^2}+\frac{1}{u^2}\right) \right], &&
\end{flalign}
\endgroup

\begingroup
\small
\allowdisplaybreaks
\begin{flalign}
 \Delta F_{bb}(s,t,u,Q^2)=&\frac{3 L_{\lambda} (t-u)^2 (t+u) \left[\left(2 s_{23}-t-u\right) (t+u)+4 s \left(s-Q^2 \right)\right]}{8 \lambda^5 \ s}+\frac{3 s_{23} (t-u)^2 \left(s+Q^2\right)}{\lambda^4 \ s} & \notag \\
 &+\frac{L_{\lambda}}{8 \lambda^3 \ s} \Bigg[ \frac{\left(t^2-u^2\right)^2+4 s \left(t^2+u^2\right) (2 s+t+u)}{\text{ds}}+4 s \left(t^2+u^2\right)\notag \\
 &\quad +\left(2 s_{23}-t-u\right) (t-u)^2 \Bigg] -\frac{4 \left(s s_{23}+t u\right)-3 (t-u)^2}{4 \lambda^2 \ s} \notag \\
 &- \frac{L_{\lambda}}{8 \lambda \ s} \left[ \frac{12 s^2+3 t^2+2 t u+3 u^2+10 s \left(t+u \right)}{\text{ds}}+2 \left(2 s-s_{23}+Q^2\right) \right]+\frac{5}{4 s}.
\end{flalign}
\endgroup

\noindent The diagrams involved in $\Delta F_{aa}(s,t,u,Q^2)$ function also contribute to the $\delta(s_{23})$ terms  in $\Delta V^{(3)}_{q\overline{q}}(s,t,u,Q^2)$. The following interference only contributes to $Z$ production:

\begingroup
\small
\allowdisplaybreaks
\begin{flalign}
 \Delta F_{ab}(s,t,u,Q^2)=&-\frac{3 L_{\lambda} s (t-u)^2 (2 s+t+u) Q^2}{2 \lambda^5 \ t}-\frac{3 (t-u)^2 Q^2 \left[2 s_{23}^2-\left(s+s_{23}\right) (t+u)\right]}{2 \lambda^4 \ t u} \notag & \\
 &+\frac{L_{\lambda} Q^2 \left[s \left(2 s-4 s_{23}+3 t-u\right)+4 \left(s_{23}-t\right) \left(s_{23}-u\right)\right]}{2 \lambda^3 \ t} \notag \\
 &+\frac{t \left(s+3 t-2 u \right)-Q^2 \left(s+2 t-u \right)}{\lambda^2 \ t} + \langle t \Leftrightarrow u \rangle.
\end{flalign}
\endgroup

The following functions have also terms proportional to $1/s_{23}$, but the expression is in fact finite in the limit $s_{23} \rightarrow 0$:

\begingroup
\small
\allowdisplaybreaks
\begin{flalign}
 \Delta F_{ac}(s,t,u,Q^2)=&\left( C_F-\frac{C_A}{2} \right) \Bigg\{ -\frac{3 L_{\lambda} s^2 (t-u)^2 (t+u) Q^2}{\lambda^5 \ t u}-\frac{3 s (t-u)^2 \left(2 s_{23}-t-u\right) Q^2}{\lambda^4 \ t u} \notag & \\
 &+\frac{L_{\lambda} s Q^2}{\lambda^3} \left[s \left(\frac{7}{u}-\frac{5}{t}\right)+4 \left(1-\frac{u}{t}\right) \right]-\frac{1}{\lambda^2} \bigg[ s^2 \left(\frac{7}{u}-\frac{5}{t}\right) \notag \\
 &\quad-\left(2 s_{23}-t-u\right) \left(1+\frac{t}{2 u}-\frac{3 u}{2 t}\right)+s \left(2+5 \ \frac{s_{23}-u}{t}-\frac{7 s_{23}-5 t}{u}\right) \bigg] \notag \\
 &+\frac{L_{\lambda}}{\lambda \ t} \left[ 2 Q^2-\left(1+\frac{t}{u}\right) \frac{s^2+(s+u)^2}{s_{23}} \right] - \frac{s^2}{\text{dt}^2 u}-\frac{3 s^2}{\text{dt} \ t u}+\frac{3 Q^2}{t^2} \notag \\
 &-\frac{2 L_{st}}{t} \left[ \frac{1}{s_{23}} \left(\frac{s^2}{u}+s+\frac{u}{2}\right)-\frac{s-s_{23}+u}{t}-1 \right]-\frac{3}{2 t}+\frac{1}{2 u} \Bigg\},
\end{flalign}
\endgroup

\begingroup
\small
\allowdisplaybreaks
\begin{flalign}
 \Delta F_{bc}(s,t,u,Q^2)=&\left( C_F-\frac{C_A}{2} \right) \Bigg\{ -\frac{6 L_{\lambda} s \ s_{23} (t-u)^2 Q^2}{\lambda^5 t}-\frac{3 (t-u)^2 \left(2 s_{23}-t-u\right)}{2 \lambda^4} \left(\frac{u}{t}+1 \right) \notag & \\
 &-\frac{L_{\lambda}}{\lambda^3} \left[ \frac{\left(2 s_{23}-t-u\right) \left(t^2-u^2\right)}{2 s}-\frac{u^2}{t} \left(\frac{s}{2}-Q^2 \right) + t \left(\frac{3 s}{2}-Q^2\right)-3 s u \right] \notag \\
 &-\frac{1}{\lambda^2} \left[ \frac{t^2-u^2}{s}+\frac{u \left(u-3 s_{23}\right)}{t}+s_{23}+t \right]-\frac{L_{\lambda}}{2 \lambda \ \text{ds}} \left[\frac{t^2}{s}+(2 s+u)(6-\frac{u}{s})\right] \notag \\
 &-\frac{L_{\lambda}}{\lambda} \left[ \frac{3 s_{23}^2+\left(s_{23}-u\right)^2}{s t}+\frac{3 t+2 u-6 s_{23}}{s}-\frac{s_{23}-u}{t}+\frac{3 s}{2 t} \right] \notag \\
 &-\frac{2}{\text{dt}}-\frac{\left(L_{st}+L_{\lambda t}\right)}{\text{ds} \ s} \left(\frac{2 s_{23}^2}{t}-2 s_{23}+t \right) +\frac{4 s_{23} Q^2}{s t^2} \notag \\ 
 &-L_{st} \left( \frac{u}{s t}+\frac{1}{s}+\frac{2}{t} \right)-\frac{2 u}{s t}-\frac{1}{2 t}+\frac{1}{s} \Bigg\},
\end{flalign}
\endgroup

\begingroup
\small
\begin{flalign}
 \Delta F_{ad}(s,t,u,Q^2)=\Delta F_{ac}(s,u,t,Q^2), &&
\end{flalign}
\endgroup

\begingroup
\small
\begin{flalign}
 \Delta F_{bd}(s,t,u,Q^2)=\Delta F_{bc}(s,u,t,Q^2), &&
\end{flalign}
\endgroup


\begingroup
\small
\allowdisplaybreaks
\begin{flalign}
 \Delta F_{cc}(s,t,u,Q^2)=&\frac{L_{\lambda} s \left(t^2-u^2\right)}{2 \lambda^3 t}+\frac{(t-u) \left(2 s_{23}-t-u\right)}{2 \lambda^2 t}-\frac{L_{\lambda} \left(3 s+2 u \right)}{2 \lambda t} \notag & \\
 &+\frac{\left(4+L_F-L_{s_{23}}\right) s}{2 \text{dt}^2}+\frac{\left(L_F-L_{s_{23}}\right) (2 s+u)}{\text{dt} \  t}+\frac{8 s-t+5 u}{2 \text{dt} \ t} \notag \\
 &+\frac{Q^2-t}{t^2} \left[ L_{st}+2 \left(L_F-L_{s_{23}}-L_{st}\right) \left(1+\frac{s_{23}}{s}\right)+\frac{4 s_{23}}{s}+\frac{5}{2} \right] \notag \\
 &+\frac{\left(1+L_F-L_{s_{23}}-L_{st}\right) \left(4 s_{23}-2 u-t\right)+2 s_{23}}{2 s t},
\end{flalign}
\endgroup

\begingroup
\small
\begin{flalign}
 \Delta F_{dd}(s,t,u,Q^2)=\Delta F_{cc}(s,u,t,Q^2). &&
\end{flalign}
\endgroup

Finally, the interference between the $F_c$ and $F_d$ diagrams yield different results depending on whether the quirality of the quark and the antiquark are the same (LR) or opposite (LL). This also makes the LR contribution equal to its unpolarized counterpart, while the LL differs by a sign.

\begingroup
\small
\allowdisplaybreaks
\begin{flalign}
 \Delta F^{LL}_{cd}(s,t,u,Q^2)=&\frac{L_{\lambda} s (t-u)^2}{2 \lambda^3 \ t}-\frac{(t-u)^2}{2 \lambda^2 \ u} \left(1-\frac{s_{23}}{t}\right)-\frac{L_{\lambda} \left(2 s-t-u \right)}{2 \lambda \ \text{ds}} - \frac{L_{\lambda}}{2 \lambda} \left( \frac{3 s+2 s_{23}}{t}-1 \right) \notag & \\
 &-\frac{\left(L_{su}-L_{\lambda t}-2 L_{tu}\right) \left[2 s \left(s+u \right)+u^2\right]}{2 \text{ds} \ \text{dst} \ t}+\frac{\left(L_{su}-L_{\lambda t}-2 L_{tu}\right)}{2 \text{dst}} \left(\frac{2s+u}{t}+1 \right) \notag \\
 &+\frac{\left(L_{\lambda t}+L_{tu}\right)}{2 \text{ds}} \left(\frac{u}{t}+1 \right)-\frac{\left(L_{st}-L_{tu}\right)}{\text{ds}} \left[ \frac{2 s \left(Q^2+s_{23}\right)}{t u}+\frac{t}{u}+\frac{u}{2 t}+\frac{1}{2} \right] \notag \\
 &-\frac{s}{\text{dt} \ u}-\frac{L_{tu} \left(s+s_{23}\right)-(u-s)}{2 t u}+\frac{L_{st}+L_{\lambda t}}{2 t} + \langle t \Leftrightarrow u \rangle,
\end{flalign}
\endgroup

\begingroup
\small
\allowdisplaybreaks
\begin{flalign}
 \Delta F^{LR}_{cd}(s,t,u,Q^2)=&\frac{L_{\lambda t}+2 L_{tu}-L_{su}}{2 t} \left( \frac{s+s_{23}-u}{\text{dst}}-1 \right)+\frac{L_{st}-L_{tu}}{t} \left( \frac{s^2}{\text{dst} \ \text{dsu}}-\frac{s+s_{23}}{u} \right) \notag & \\
 &-\frac{s^2}{\text{dst} \ t} \left[ \frac{\left( L_{st}+L_{\lambda t} \right)}{\text{ds}} - \frac{ \left(L_{su}-L_{tu} \right)}{u} \right] + \langle t \Leftrightarrow u \rangle.
\end{flalign}
\endgroup

\subsubsection{Unpolarized $q \overline{q} \rightarrow V X$}

Due to helicity conservation, most of the unpolarized results differ from the polarized one only by a sign, e.g. $G_{q\overline{q}}(s,t,u,Q^2)=-\Delta G_{q\overline{q}}(s,t,u,Q^2)$, so we list only the functions in which this is not true:

\begingroup
\small
\allowdisplaybreaks
\begin{flalign}
 F_{cc}(s,t,u,Q^2)=&-\Delta F_{cc}(s,t,u,Q^2) + \frac{3 s}{2 \text{dt}^2}+\frac{4 s+3 u}{\text{dt} \ t}-\left(L_F-L_{s_{23}}-L_{st}\right) \bigg[\frac{1}{2 s}\left(\frac{2 s_{23} Q^2}{t^2}-\frac{u}{t}\right)^2 \notag & \\
 & \quad +\frac{\left(2 s_{23}-t-u\right)^2-4 s_{23} \left(Q^2-t\right)}{2 \ s \ t^2}-\frac{4 s_{23}-2 u-t}{2 \ s \ t}\bigg] \notag \\
 &-\frac{3}{2 s} \left(\frac{2 s_{23} Q^2}{t^2}-\frac{u}{t}\right)^2 -\frac{10 s_{23}^2-8 s_{23} u-3 u^2}{2 s t^2} +\frac{4 s_{23}-u}{s t}+\frac{3 s+2 s_{23}+4 u}{t^2}-\frac{1}{2 s},
\end{flalign}
\endgroup

\begingroup
\small
\begin{flalign}
 F_{dd}(s,t,u,Q^2)=F_{cc}(s,u,t,Q^2), &&
\end{flalign}
\endgroup

\begingroup
\small
\allowdisplaybreaks
\begin{flalign}
 F^{LR}_{cd}(s,t,u,Q^2)=\Delta F^{LR}_{cd}(s,t,u,Q^2).&&
\end{flalign}
\endgroup

\subsection{Quark-gluon Compton scattering}

In this section we present the functions corresponding to the quark-gluon Compton processes in both the polarized and unpolarized case. The diagrams are obtained from Figs. \ref{fig_FeynVirt} and \ref{fig_FeynqqGG} by suitable crossing.

\subsubsection{Polarized $q g \rightarrow V X$}

We start with the $\delta(s_{23})$ contributions from virtual and real contributions

\begingroup
\small
\begin{flalign}
 \Delta V^{(1)}_{qg}(s,t,u,Q^2)=\Delta B_{qg}(s,t,u,Q^2) \left[ \frac{1}{3} \left(L_F-L_{R}\right) \right], &&
\end{flalign}
\endgroup

\begingroup
\small
\begin{flalign}
 \Delta V^{(2)}_{qg}(s,t,u,Q^2)=\frac{(s-t)}{s+t} \left[ 1+\frac{L_u Q^2}{s+t} \right], &&
\end{flalign}
\endgroup

\begingroup
\small
\begin{flalign}
 \Delta V^{(3)}_{qg}(s,t,u,Q^2)=&\Delta B_{qg}(s,t,u,Q^2) \Bigg\{ C_F \bigg[ L_A^2-L_u^2-L_A \left(2 L_F+\frac{3}{2}\right)  \notag & \\
 &\quad -L_F \left(\frac{3}{2}-2 L_u\right)-5-\frac{1}{3} \pi^2 \bigg] + C_A \bigg[2 L_A^2+L_u^2-2 L_A \left(L_F-L_s+L_t+L_u\right) \notag \\
 &\quad -L_F \left(\frac{11}{6}-2 L_t\right)+\frac{11}{6} L_{R}+L_{1t}-L_{2t}-L_s \left(L_t+L_u\right)+L_t L_u+\frac{1}{2}+\frac{\pi^2}{2} \bigg] \Bigg\} \notag \\
 &+\frac{L_s \ s}{(t+u)} \left[ C_F \left(1+\frac{4 u}{s}\right)+\frac{C_F \ u}{(t+u)}+C_A \right] - \frac{C_F \ t}{(t+u)} \notag \\
 &-\frac{L_t \ t}{(s+u)} \left[ C_F \left(1+\frac{4 u}{t}\right)+\frac{C_F \ u}{(s+u)}+C_A \right] + \frac{C_F \ s}{(s+u)} \notag \\
 &-\left(C_F-C_A\right) \frac{s^2-t^2}{2 s t}-\frac{\left(2 C_F-C_A\right) (t+2 u)}{s} \left[ L_{1u}-L_{2u}\right] \notag \\
 &+\frac{\left(2 C_F-C_A\right) (s+2 u)}{t} \left[ L_{1t}+L_{1u}-L_t L_u+\frac{\pi^2}{2}  \right],
\end{flalign}
\endgroup

\noindent where $\Delta B_{qg}(s,t,u,Q^2)$ is the Born function defined in Eq.(\ref{eq_Bornqg}). Here again the $\Delta V^{(2)}_{qg}(s,t,u,Q^2)$ terms arise from the triangle quark loops diagrams, which only contribute to $Z$ production.

\begingroup
\small
\allowdisplaybreaks
\begin{flalign}
 \Delta G_{qg}(s,t,u,Q^2)=&\frac{(s-t) \left(s+t+2 u \right)}{s t} \bigg\{ \left(2 C_F+4 C_A\right) \left( \frac{L_{s_{23}}}{s_{23}}\right)_{A+} -\left( \frac{1}{s_{23}}\right)_{A+} \notag & \\
 &\quad \times \bigg[C_F \left(\frac{3}{2}+2 L_F-L_{\text{su}}-2 L_{\text{tu}}-L_{\lambda} \frac{t+u}{\lambda }\right) \notag \\
 &\quad \quad-C_A \left(2 L_{\text{stu}}-2 L_F+\frac{L_{\text{st}}-L_{\text{su}}}{2}-L_{\text{tu}}\right) \bigg] \bigg\} \notag \\
 &-\frac{2 C_F-C_A}{2 s_{23}} \Bigg[ \frac{L_{\text{su}} \left(s+2 u \right)}{t}-\frac{2 L_{\text{tu}} \left(t+2 u \right)}{s}+\frac{L_{\lambda} \left(s+2 u \right)}{\lambda} \left(\frac{u}{t}+1 \right) \Bigg] \notag \\
 &+\frac{3 C_F L_{\lambda} s (t-u)^2 (2 Q^2-t-u)}{4 \lambda^5} \left(\frac{u}{t}+1 \right) -\frac{3 C_F s (t-u)^2}{2 \lambda^4} \left(\frac{u}{t}+1 \right) \notag \\
 &-\frac{L_{\lambda}}{\lambda^3} \bigg\{ \frac{C_F}{2} \bigg[ \frac{\left(t^2-u^2\right) \left(t+u-2 Q^2\right)}{2 s}+u \ \frac{(t-u)^2-s u-Q^2 \left(s+2 u \right)}{t} +(t-u)^2 \notag \\
 & \quad -s t+Q^2 \left(3 s-6 t+8 u \right) \bigg] +C_A \left(1-\frac{u}{t} \right) \bigg[s^2-\frac{\left(2 s_{23}-t-u\right) \left(s+Q^2+s_{23}\right)}{4}\bigg] \bigg\} \notag \\
 &+\frac{1}{\lambda^2} \bigg\{ \frac{C_F}{2} \left[ s \left(3-\frac{u}{t}\right)-(t-u) \left(\frac{3 Q^2-s_{23}}{s}+\frac{4 s_{23}-7 u}{2 t}+\frac{5}{2}\right) \right] \notag \\
 &\quad +C_A \left(s+s_{23}\right) \left(1-\frac{u}{t}\right)\bigg\} + \frac{L_{\lambda}}{2 \lambda} \bigg[ C_F \left(\frac{9 s-20 Q^2+6 u}{2 t}-\frac{2 Q^2-9 t-11 u}{2 s}+3\right) \notag \\
 &\quad +C_A \left( \frac{6 Q^4-5 u Q^2+u^2}{s t}-\frac{12 s+9 u+14 Q^2}{2 t}+\frac{u-3 Q^2}{s}-\frac{3}{2} \right) \bigg] \notag \\
 &-\frac{C_F \left(L_{su}-2 L_{tu}-L_{\lambda t}\right) s_{23} \left(s_{23}-2 u\right)}{\text{dst} \ s t}+\frac{C_F}{\text{dq}} \Bigg\{ Q^2 \left(\frac{4 u}{t}-1\right)+s-3 u \notag \\
 &\quad + 2 \left(L_F-L_{s_{23}}-L_{tu}\right) \left[Q^2 \left(\frac{4 s+6 u-2 Q^2}{t}+1\right)+s+2 t-5 u\right] \Bigg\} \notag \\
 &+ \frac{\left(5 C_F+7 C_A\right) s}{\text{dt}^2}+\frac{2 \left(C_F+2 C_A\right) \left(L_F-L_{s_{23}}\right) s}{\text{dt}^2} \notag \\ 
 &+\frac{C_F \left[3 s-8 t+2 u-\left(L_F-L_{s_{23}}\right) (s+2t+2u)\right]}{\text{dt} \ t} +\frac{C_A}{\text{dt}} \bigg[ \frac{\left(L_{st}+L_{\lambda t}\right)}{2} \notag \\
 & \quad \times \left(\frac{1}{s}-\frac{3}{t}-\frac{2 u}{s t}\right) (s+t) +\left(L_{s_{23}}+L_{stu}-L_{tu}\right) (s-t) \left(\frac{1}{s}-\frac{1}{t}-\frac{2 u}{s t}\right)\bigg] \notag \\
 &+\frac{C_A \left[15 s+t+10 u+2 \left(L_F-L_{s_{23}}\right) (5 s+4 u)\right]}{\text{dt} \ t} +\frac{\left(2 C_F-C_A\right)}{\text{du}} \left(\frac{s}{t}+1\right) \notag \\
 &+\frac{C_F \left(1+L_F-L_{s_{23}}-L_{su}\right) s_{23} \left(s-s_{23}+t\right)}{s u^2} \notag \\
 &-\frac{C_F \left[s+t+\left(L_F-L_{s_{23}}-L_{su}\right) \left(s-4 s_{23}+2 t\right)\right]}{s u}+\frac{s-s_{23}+u}{t^2} \bigg\{ \left(C_F+2 C_A\right) L_{st} \notag \\
 & \quad +\left(L_F-L_{s_{23}}-L_{st}\right) \left[C_F+8 C_A \left(1+\frac{s_{23}}{s}\right)\right]+\frac{5}{2} C_F+C_A \left(9+\frac{12 s_{23}}{s}\right) \bigg\} \notag \\
 &-2 \left[C_F \left(\frac{3}{t}-\frac{2}{s}\right)-2 C_A \left(\frac{2 s_{23}}{s t}-\frac{u}{s t} \right) \right] \left(L_F-L_{s_{23}}\right) -\frac{s_{23}-2 u}{2 s t} \bigg[2 C_F L_{su} \notag \\
 &\quad +C_A \left(2 L_{s_{23}}+2 L_{stu}-L_{su}\right) \bigg]-\frac{L_{\lambda t}}{t} \left[\frac{C_F \left(3 s-s_{23}+2 u\right)}{s}-2 C_A\right] \notag \\
 &+\frac{2 C_F L_{tu} \left(3 s+s_{23}-2 u\right)}{s t}-\frac{C_A L_{st} \left(2 s+9 s_{23}-4 u\right)}{s t} \notag \\
 &-\frac{17 C_F}{4 t}+\frac{2 C_A \left(4 s_{23}-u\right)}{s t} - \frac{C_F}{s} \left[2+5 \left(2 L_F-2 L_{s_{23}}-L_{su}\right)+3 L_{\lambda t}\right] \notag \\
 &-\frac{C_A}{s} \left[1+2 \left(3 L_F-4 L_{s_{23}}-L_{stu}-L_{\lambda t}\right)- \frac{9}{2} L_{st}+L_{su}-L_{tu}\right].
\end{flalign}
\endgroup

\noindent Here the $1/s_{23}$ terms without the corresponding `+' distributions are actually finite in the limit $s_{23} \rightarrow 0$.

\subsubsection{Unpolarized $q g \rightarrow V X$}

The $q g \rightarrow V X$ process is the one that presents more differences between the expressions of the polarized and unpolarized cases:

\begingroup
\small
\begin{flalign}
 V^{(1)}_{qg}(s,t,u,Q^2)=B_{qg}(s,t,u,Q^2) \left[ \frac{1}{3} \left(L_F-L_{R}\right) \right], &&
\end{flalign}
\endgroup

\begingroup
\small
\begin{flalign}
 V^{(2)}_{qg}(s,t,u,Q^2)=-\frac{(Q^2+u)}{(Q^2-u)} \left[ 1+\frac{L_u Q^2}{(Q^2-u)} \right], &&
\end{flalign}
\endgroup

\begingroup
\small
\begin{flalign}
 V^{(3)}_{qg}(s,t,u,Q^2)=& B_{qg}(s,t,u,Q^2) \Bigg\{ C_F \bigg[ L_A^2-L_u^2-L_A \left(2 L_F+\frac{3}{2}\right)  \notag & \\
 &\quad -L_F \left(\frac{3}{2}-2 L_u\right)-5-\frac{1}{3} \pi^2 \bigg] + C_A \bigg[2 L_A^2+L_u^2-2 L_A \left(L_F-L_s+L_t+L_u\right) \notag \\
 &\quad -L_F \left(\frac{11}{6}-2 L_t\right)+\frac{11}{6} L_{R}+L_{1t}-L_{2t}-L_s \left(L_t+L_u\right)+L_t L_u+\frac{1}{2}+\frac{\pi^2}{2} \bigg] \Bigg\} \notag \\
 &-\frac{L_s \ s}{(t+u)} \left[ C_F \left(1+\frac{4 u}{s}\right)+\frac{C_F \ u}{(t+u)}+C_A \right] + \frac{C_F \ t}{(t+u)} \notag \\
 &-\frac{L_t \ t}{(s+u)} \left[ C_F \left(1+\frac{4 u}{t}\right)+\frac{C_F \ u}{(s+u)}+C_A \right] + \frac{C_F \ s}{(s+u)} \notag \\
 &+\frac{\left(2 C_F-C_A \right)}{(s+t)} \left[ \frac{Q^2 \left(s^2+t^2\right)}{s t}-\frac{2 L_u \ u \left(2
   Q^2-u\right)}{(s+t)} \right] \notag \\
 &- C_F \left( 2+\frac{3 t+4 u}{2 s}+\frac{3 s+4 u}{2 t}+\frac{u^2}{s t} \right) + C_A \left( \frac{t+2 u}{2 s}+\frac{s+2 u}{2 t}+\frac{u^2}{s t} \right)\notag \\
 &-\left(2 C_F-C_A \right) \bigg[ \frac{u^2+(s+u)^2}{s t} \left( L_{1t}+L_{1u}-L_t L_u+\frac{\pi^2}{2} \right) \notag \\
 & \quad + \frac{u^2+(u+t)^2}{s t} \left( L_{1u}-L_{2u} \right) \bigg],
\end{flalign}
\endgroup

\noindent where $B_{qg}(s,t,u,Q^2)$ is the unpolarized Born function defined in Eq.(\ref{eq_Bornqg}). We note that the virtual contributions presented in Eq.(A4) of Ref.\cite{Gonsalves:1989ar}, in our case included within $V^{(3)}_{qg}(s,t,u,Q^2)$, 
are missing a minus sign in the finite terms not multiplied by the Born function.

\begingroup
\small
\allowdisplaybreaks
\begin{flalign}
 G_{qg}(s,t,u,Q^2)=&-\left[\frac{t}{s}+ \frac{s}{t} +\frac{2 u \left(s+t+u \right)}{s t} \right] \bigg\{ \left(2 C_F+4 C_A\right) \left( \frac{L_{s_{23}}}{s_{23}}\right)_{A+} -\left( \frac{1}{s_{23}}\right)_{A+} \notag & \\
 &\quad \times \bigg[C_F \left(\frac{3}{2}+2 L_F-L_{\text{su}}-2 L_{\text{tu}}-L_{\lambda} \frac{t+u}{\lambda}\right) \notag \\
 &\quad \quad -C_A \left(2 L_{\text{stu}}-2 L_F+\frac{L_{\text{st}}-L_{\text{su}}}{2}-L_{\text{tu}}\right) \bigg] \bigg\} \notag \\
 &+\frac{2 C_F-C_A}{2 s_{23}} \Bigg[ \frac{L_{\text{su}} \left(s+2 u \right)}{t}+\frac{2 L_{\text{tu}} \left(t+2 u \right)}{s}+\frac{L_{\lambda} \left(s+2 u \right)}{\lambda} \left(\frac{u}{t}+1 \right) \Bigg] \notag \\
 &-\frac{3 C_F L_{\lambda} s (t-u)^2 (2 Q^2-t-u)}{4 \lambda^5} \left(\frac{u}{t}+1 \right) +\frac{3 C_F s (t-u)^2}{2 \lambda^4} \left(\frac{u}{t}+1 \right) \notag \\
 &+\frac{L_{\lambda}}{\lambda^3} \bigg\{\frac{C_F}{2} \bigg[ \frac{\left(t^2-u^2\right) \left(t+u-2 Q^2\right)}{2 s} +u \ \frac{(t-u)^2-s u-Q^2 \left(s+2 u \right)}{t}+(t-u)^2 \notag \\
 & \quad -s t+Q^2 \left(3 s-6 t+8 u \right) \bigg] + C_A \left(1-\frac{u}{t} \right) \bigg[s^2-\frac{\left(2 s_{23}-t-u\right) \left(s+Q^2+s_{23}\right)}{4}\bigg] \bigg\} \notag \\
 &-\frac{1}{\lambda^2} \bigg\{ \frac{C_F}{2} \left[ s \left(3-\frac{u}{t}\right)-(t-u) \left(\frac{3 Q^2-s_{23}}{s}+\frac{4 s_{23}-7 u}{2 t}+\frac{5}{2}\right) \right] \notag \\
 &\quad +C_A \frac{\left(s+s_{23}\right) \left(t-u\right)}{t}\bigg\}-\frac{L_{\lambda}}{2 \lambda} \Bigg\{ C_F \Big[8 \ \frac{2 Q^2 \left(s_{23}-Q^2\right)-u^2}{s t} -\frac{2 Q^2+7 t+21 u}{2 s} \notag \\
 &\quad-\frac{7 s-28 Q^2+2 u}{2 t}-1 \Big] +C_A \bigg[\frac{6 Q^4-5 u Q^2+u^2}{s t}-\frac{12 s+9 u+14 Q^2}{2 t} \notag \\
 &\quad \quad +\frac{u-3 Q^2}{s}-\frac{3}{2} \bigg] \Bigg\} -\frac{C_F \left(L_{su}-2 L_{tu}-L_{\lambda t}\right) \left[(s_{23}-u)^2+u^2\right]}{\text{dst} \ s t} \notag \\
 &+\frac{C_F}{\text{dq}} \bigg\{ \frac{4 (t-u) u}{s} \left( \frac{Q^2}{t}-1 \right) - u \left( \frac{4 Q^2}{t}+1 \right) -s+Q^2  +4 \left(L_F-L_{s_{23}}-L_{tu}\right) \notag \\
 &\quad \times \left[ \frac{(t-u)^2+u^2}{s}+\frac{s+u}{2}+t-Q^2 \left(\frac{2 u^2}{s t}+\frac{t-2 u}{s}+\frac{s+2 u}{t}+\frac{1}{2}\right) \right] \bigg\} \notag \\
 &+\frac{2 \left[\left(C_F-C_A \right) \left(4+L_F-L_{s_{23}}\right)+C_A \right] s t}{\text{dt}^3} \notag \\
 &-\frac{2 \left(L_F-L_{s_{23}}\right) \left[C_F \left(t+u \right)+C_A \left(3 s+2 u \right)\right]}{\text{dt}^2}+\frac{C_F \left(3 s-8 t-4 u \right)}{\text{dt}^2} \notag \\
 &-\frac{C_A \left(9 s-4 t+4 u \right)}{\text{dt}^2}+\frac{C_F}{\text{dt}} \bigg[4 \left(L_F-L_{s_{23}}\right) \left(\frac{u^2}{s t}+\frac{t-2 u}{2 s}+3 \ \frac{s+2 u}{4 t}-1\right) +\frac{u^2}{s t} \notag \\
 &\quad +\frac{3 t-2 u}{s}-\frac{s-2 u}{t}-4\bigg] -\frac{C_A}{\text{dt}} \bigg[ 4 \left(L_F-L_{s_{23}}\right) \left(\frac{u^2}{s t}+\frac{t-2 u}{2 s}+3 \ \frac{s+u}{t}-2\right) \notag \\
 &\quad -\left(L_{st}+L_{\lambda t}\right) \left(\frac{u^2}{s t}+\frac{5 s+6 u}{2 t}+\frac{t-2 u}{2 s}-1\right) +\frac{3 s+2 u}{t}-7 \notag \\
 &\quad - \left(L_{s_{23}}+L_{stu}-L_{tu}\right) \frac{(s-t+u)^2+u^2}{s t} \bigg] -\frac{2 C_F \left[s-\left(L_F-L_{s_{23}}\right) t\right]}{\text{du} \ t} \notag \\
 &+\frac{C_A \left(s-t \right)}{\text{du} \ t} +\frac{C_F \left(1+L_F-L_{s_{23}}-L_{su}\right) s_{23} \left(s-s_{23}+t\right)}{s u^2} \notag \\
 &+\frac{C_F}{u} \left[ 2 \left(L_F-L_{s_{23}}-L_{su}\right) \left(\frac{2 s_{23}-t}{s}-\frac{1}{2}\right)-\frac{t}{s}-1 \right] \notag \\
 &-\frac{4 C_A \left(3+L_F-L_{s_{23}}-L_{st}\right) s_{23} \left(s-s_{23}+u\right)}{s t^3} \left[ \frac{s_{23} \left(s-s_{23}+u\right)}{t}+\left(2 s_{23}-u\right) \right] \notag \\
 &+\frac{s-s_{23}+u}{t^2} \bigg\{ C_F \left[\left(L_F-L_{s_{23}}\right)-\frac{1}{2}\right] +C_A \bigg[4 \left(L_F-L_{s_{23}}-L_{st}\right) \left(\frac{3 s_{23}-2 u}{s}-2\right) \notag \\
 & \quad -2 L_{st} +\frac{16 s_{23}}{s}-1\bigg] \bigg\} -\frac{2 C_A}{t^2} \left[ \frac{\left(L_F-L_{s_{23}}-L_{st}\right) \left(s+s_{23}\right)^2}{s}+\frac{6 s \ s_{23}+u^2}{s} \right] \notag \\
 &+\frac{L_{\lambda t} \left[C_F \left(5 s-7 s_{23}+6 u\right)-2 C_A \left(3 s-2 s_{23}+2 u\right)\right]}{s t} -\frac{2 C_F L_{tu} \left(s+s_{23}-2 u\right)}{s t} \notag \\
 &+\frac{\left[2 C_F \left(4 L_{tu}-L_{su}\right)+C_A \left(L_{su}-2 L_{s_{23}}-2 L_{stu}\right)\right] \left(s_{23}-2 u\right)}{2 s t} \notag \\
 &-\frac{2 \left[C_F \left(s+2 t \right)-C_A \left(5 s_{23}-4 u\right)\right] \left(L_F-L_{s_{23}}\right)}{s t} +\frac{2 C_A L_{st} \left(3 s-6 s_{23}+5 u\right)}{s t}  \notag \\
 &+\frac{C_F s+8 C_A \left(2 s_{23}-u\right)}{4 s t} -\frac{C_F}{s} \left[2+4 \left(L_F-L_{s_{23}}-L_{tu}\right)-5 \left( L_{su}+L_{\lambda t} \right) \right] \notag \\
 &-\frac{C_A}{s} \left[1+2 \left(4 L_F-6 L_{s_{23}}-2 L_{stu}+L_{\lambda t}\right)-\frac{11}{2} L_{st}+L_{su}+L_{tu}\right].
\end{flalign}
\endgroup

\subsection{Gluon-gluon fusion}

We now present the functions for the gluon-gluon fusion processes for both the polarized and unpolarized cases. The diagrams are obtained from Fig.\ref{fig_FeynqqGG} by suitable crossing.

\subsubsection{Polarized $g g \rightarrow V X$}

\begingroup
\small
\allowdisplaybreaks
\begin{flalign}
 \Delta G_{gg}(s,t,u,Q^2)=&-\frac{3 C_A L_{\lambda} (t-u)^2 t}{\lambda ^5} \left[\frac{\left(2 s_{23}-t-u\right) (t+u)}{4 s}+s-Q^2 \right] \notag & \\
 &+\frac{C_A L_{\lambda}}{\lambda^3 \text{ds}} \left\{\frac{s_{23} \left(s_{23}-t-u\right) (t-u)^2}{2 s}-t \left[ 4 s t-2 t u-s_{23} (t+u)+\frac{3}{2} (t+u)^2 \right]\right\} \notag \\
 &-\frac{L_{\lambda}}{\lambda \ \text{ds}} \left\{4 C_F \left[2 (s+t+u)-3 s_{23}\right]-\frac{C_A}{2} \left[ \frac{14 Q^4+\left(t-Q^2\right)^2}{s}+9 Q^2+3 s \right] \right\} \notag \\
 &-\frac{C_A L_{\lambda}}{\lambda} \left(1-\frac{14 Q^2-21 t-17 u}{8 s} \right) -\frac{3 C_A s_{23} (t-u)^2}{\lambda^4} \left(\frac{Q^2}{s}+1\right) \notag \\
 &+\frac{C_A}{\lambda^2} \left[\frac{t (5 u-3 t)}{2 s}+s_{23} \right] -\frac{2 C_F \left(L_{st}-L_{tu}\right) \left[s_{23}^2+\left(s-s_{23}\right)^2 \right]}{\text{dst} \ \text{dsu} \ s} \notag \\
 &+\frac{2 C_F \left(L_{st}+L_{\lambda t}\right) \left(s_{23}^2+t^2\right)}{\text{ds} \ \text{dst} \ s}-\frac{2 C_F \left(L_{su}-2 L_{tu}-L_{\lambda t}\right) (t+u)}{\text{dst} \ s} \notag \\
 &-\frac{2 C_F \left[\left(L_F-L_{s_{23}}-L_{tu}\right) (s+4 t)+2 s_{23}+2 t\right]}{\text{dq}} \notag \\
 &+\frac{2 \left(2 C_F-C_A\right) \left(L_{s_{23}}+L_{stu}-L_{tu}\right)}{\text{dt} \ s} \left( \frac{t u}{\text{du}}+t-u \right) \notag \\
 &+\frac{2 \left(2 C_F-C_A\right) \left(L_{st}+L_{\lambda t}\right)}{\text{dt}} \left( \frac{t^2}{\text{ds} \ s}+\frac{t+u}{2 s}+1 \right) +\frac{\left(2 C_F-C_A\right) \left(L_{st}+L_{\lambda t}\right)}{2 \text{ds}} \notag \\
 & \quad \times \left(\frac{2 s_{23}+5 t+u}{2 s}+1\right) -\frac{2 C_F \left(4+L_F-L_{s_{23}}-L_{\text{st}}\right) s_{23} \left(Q^2-t\right)}{s t^2} \notag \\
 &-\frac{2 C_F}{t} \left[\left(L_F-L_{s_{23}}-L_{\text{st}}\right) \left(\frac{s_{23}+u}{s}+2\right)+\frac{6 s_{23}-u}{s}+2 \right] \notag \\
 &-\frac{C_F \left[L_{\lambda t}-7 L_{st}+8 \left(L_F-2 L_{s_{23}}-L_{stu}+L_{tu}\right)\right]}{2 s} \notag \\
 &-\frac{C_A \left[17-L_{st}+8 \left(L_{s_{23}}+L_{stu}\right)-9 L_{\lambda t}\right]}{4 s} + \langle t \Leftrightarrow u \rangle.
\end{flalign}
\endgroup

\subsubsection{Unpolarized $g g \rightarrow V X$}

\begingroup
\small
\allowdisplaybreaks
\begin{flalign}
 G_{gg}(s,t,u,Q^2)=&-\Delta G_{gg}(s,t,u,Q^2)+4 \ \bigg\{ -\frac{C_F \left[s \ s_{23}-\left(1+2 L_F-2 L_{s_{23}}-2 L_{tu}\right) t^2-t u\right]}{\text{dq} \ s} \notag & \\
 &\quad +\frac{\left(2 C_F-C_A\right) \left(L_{s_{23}}+L_{stu}-L_{tu}\right) t u}{\text{dt} \ \text{du} \ s} +\frac{2 \left[C_F \left(1+L_F-L_{s_{23}}\right)+C_A\right] t}{\text{dt}^2} \notag \\
 &\quad +\frac{\left(2 C_F-C_A\right) \left(L_{s_{23}}+L_{stu}-L_{tu}\right) (t-u)}{\text{dt} \ s} \notag \\
 &\quad + \frac{2 C_F \left(1+L_F-L_{s_{23}}\right) s+C_A \left(2 s-t+u \right)}{\text{dt} \ s}-\frac{C_F s_{23} Q^2}{s t^2}+\frac{C_F \left(u-2 s_{23}\right)}{s t} \notag \\
 &\quad + \frac{C_F \left(1-2 L_F+4 L_{s_{23}}+2 L_{stu}\right)-C_A \left(2+L_{s_{23}}+L_{stu}\right)}{s} \bigg\}+ \langle t \Leftrightarrow u \rangle,
\end{flalign}
\endgroup

\noindent where in this case the $\langle t \Leftrightarrow u \rangle$ only applies to the additional terms added to $\Delta G_{gg}(s,t,u,Q^2)$, which itself is already symmetric in $t$ and $u$.

\subsection{Quark-quark scattering}

Finally, we present the functions corresponding to the quark-quark scattering processes for both the polarized and unpolarized cases. They are obtained from the diagrams of Fig.\ref{fig_Feynqq}. Most of them are related to the quark-antiquark results.

\subsubsection{Polarized $q q \rightarrow V X$}

\begingroup
\small
\begin{flalign}
 \Delta H_{aa}(s,t,u,Q^2)=\Delta H_{cc}(s,t,u,Q^2)=\Delta F_{cc}(s,t,u,Q^2), &&
\end{flalign}
\endgroup

\begingroup
\small
\begin{flalign}
 \Delta H_{bb}(s,t,u,Q^2)=\Delta H_{dd}(s,t,u,Q^2)=\Delta F_{dd}(s,t,u,Q^2). &&
\end{flalign}
\endgroup

\noindent For the following functions we use the fact that changing an antiquark in the diagrams in Fig.\ref{fig_Feynqqbqqb} with a quark in Fig.\ref{fig_Feynqq} also changes its coupling to the boson $V$ by the replacement $L \leftrightarrow -R^{\dagger}$, which accounts for a change in sign in the contributions. This relative sign was overlooked in the results presented in Ref.\cite{Chang:1997ik}.

\begingroup
\small
\begin{flalign}
 \Delta H^{LL}_{ab}(s,t,u,Q^2)=\Delta H^{LL}_{cd}(s,t,u,Q^2)=-\Delta F^{LR}_{cd}(s,t,u,Q^2), &&
\end{flalign}
\endgroup

\begingroup
\small
\begin{flalign}
 \Delta H^{LR}_{ab}(s,t,u,Q^2)=\Delta H^{LR}_{cd}(s,t,u,Q^2)=-\Delta F^{LL}_{cd}(s,t,u,Q^2). &&
\end{flalign}
\endgroup

\noindent The last contributions are exclusive to the quark-quark process, and they are equal to their unpolarized counterparts:

\begingroup
\small
\allowdisplaybreaks
\begin{flalign}
 \Delta H_{ac}(s,t,u,Q^2)=\Delta H_{bd}(s,u,t,Q^2)=&\left( C_F-\frac{C_A}{2} \right) \Bigg\{ \frac{4 L_{st} Q^2}{t^2} \notag & \\
 &\quad + \left(L_{st}+L_{\lambda t}\right) \left[\frac{4 s^2}{\text{ds} \ \text{dt} \ t}-\frac{2 \left(s_{23}-u\right)}{\text{dt} \ t}-\frac{2}{t}\right] \Bigg\},
\end{flalign}
\endgroup

\begingroup
\small
\allowdisplaybreaks
\begin{flalign}
 \Delta H_{ad}(s,t,u,Q^2)=\Delta H_{bc}(s,u,t,Q^2)=&\left( C_F-\frac{C_A}{2} \right) \Bigg\{ \frac{4 s_{23} \left(s-s_{23}+u\right)}{s t^2}+\frac{2 \left( 2 s_{23}-u \right)}{s t} -\frac{2}{s} \notag & \\
 &\quad-\frac{\left(L_{s_{23}}-L_{st}+L_{stu}\right)}{s \ t \ u} \left[s_{23}^2+\left(s_{23}-t-u\right)^2\right] \Bigg\} + \langle t \Leftrightarrow u \rangle.
\end{flalign}
\endgroup

\subsubsection{Unpolarized $q q \rightarrow V X$}

For the unpolarized case we have the same relations to the quark-antiquark process:

\begingroup
\small
\begin{flalign}
 H_{aa}(s,t,u,Q^2)=H_{cc}(s,t,u,Q^2)=F_{cc}(s,t,u,Q^2), &&
\end{flalign}
\endgroup

\begingroup
\small
\begin{flalign}
 H_{bb}(s,t,u,Q^2)=H_{dd}(s,t,u,Q^2)=F_{dd}(s,t,u,Q^2), &&
\end{flalign}
\endgroup

\begingroup
\small
\begin{flalign}
 H^{LL}_{ab}(s,t,u,Q^2)=H^{LL}_{cd}(s,t,u,Q^2)=-F^{LR}_{cd}(s,t,u,Q^2), &&
\end{flalign}
\endgroup

\begingroup
\small
\begin{flalign}
 H^{LR}_{ab}(s,t,u,Q^2)=H^{LR}_{cd}(s,t,u,Q^2)=-F^{LL}_{cd}(s,t,u,Q^2). &&
\end{flalign}
\endgroup

As it was already mentioned, the contributions exclusive to the quark-quark process are equal to their polarized counterparts:

\begingroup
\small
\allowdisplaybreaks
\begin{flalign}
 H_{ac}(s,t,u,Q^2)= H_{bd}(s,u,t,Q^2)= \Delta H_{ac}(s,t,u,Q^2) &&
\end{flalign}
\endgroup

\begingroup
\small
\allowdisplaybreaks
\begin{flalign}
 H_{ad}(s,t,u,Q^2)=H_{bc}(s,u,t,Q^2)=\Delta H_{ad}(s,t,u,Q^2) &&
\end{flalign}
\endgroup


\end{document}